\documentclass[journal,11pt, onecolumn]{IEEEtran}

\ifCLASSINFOpdf
\else
   \usepackage[dvips]{graphicx}
\fi
\usepackage{url}
\usepackage{algorithmicx}
\usepackage{algpseudocode}
\usepackage{algorithm}
\algdef{SE}[DOWHILE]{Do}{doWhile}{\algorithmicdo}[1]{\algorithmicwhile\ #1}%
\usepackage{amsmath}
\usepackage{dsfont, cuted}
\usepackage{amsfonts}
\usepackage{amsthm}
\usepackage[utf8]{inputenc}

\usepackage{subfigure}
\usepackage{flushend}
\usepackage{breqn}
\usepackage{multirow}
\usepackage{multicol}
\usepackage{booktabs}
\usepackage{fullwidth}
\usepackage[noadjust]{cite}
\usepackage{setspace}
\usepackage{hyperref}
\usepackage{bmpsize}
\usepackage{flushend}
\usepackage{xcolor}
\usepackage{graphicx}
\newtheorem{thm}{Theorem}

\usepackage{multirow,tabularx}
\newcolumntype{C}{>{\centering\arraybackslash}X}
\newcolumntype{L}{>{\raggedright \arraybackslash}X}
\newcolumntype{R}{>{\raggedleft \arraybackslash}X}

\begin{document}

\title{Detection of Line Artefacts in Lung Ultrasound Images of COVID-19 Patients via Non-Convex Regularization}

\author{Oktay~Karakuş,~\IEEEmembership{Member,~IEEE,} Nantheera Anantrasirichai,~\IEEEmembership{Member,~IEEE,} 
Amazigh~Aguersif, Stein~Silva,
Adrian~Basarab,~\IEEEmembership{Senior Member,~IEEE,} Alin~Achim,~\IEEEmembership{Senior Member,~IEEE}
        \thanks{This work was supported in part by the UK Engineering and Physical Sciences Research Council (EPSRC) under grant EP/R009260/1, in part by an EPSRC Impact Acceleration Award (IAA) from the University of Bristol, and in part by a Leverhulme Trust Research Fellowship to Achim (INFHER).}
        \thanks{Oktay Karakuş, Nantheera Anantrasirichai and Alin Achim are with the Visual Information Laboratory, University of Bristol, Bristol BS1 5DD, U.K. (e-mail: o.karakus@bristol.ac.uk; n.anantrasirichai@bristol.ac.uk; alin.achim@bristol.ac.uk)}
        \thanks{
        Amazigh Aguersif and Stein Silva are with Service de Réanimation, CHU Purpan, 31300, Toulouse, France. (e-mail: amazighaguersif@gmail.com; silvastein@me.com)}
        \thanks{
        Adrian Basarab is with IRIT, University of Toulouse, CNRS UMR 5505, Toulouse, France. (e-mail: adrian.basarab@irit.fr)}
}

\maketitle

\begin{abstract}
In this paper, we present a novel method for line artefacts quantification in lung ultrasound (LUS) images of COVID-19 patients. We formulate this as a non-convex regularisation problem involving a sparsity-enforcing, Cauchy-based penalty function, and the inverse Radon transform. We employ a simple local maxima detection technique in the Radon transform domain, associated with known clinical definitions of line artefacts. Despite being non-convex, the proposed technique is guaranteed to convergence through our proposed Cauchy proximal splitting (CPS) method, and accurately identifies both horizontal and vertical line artefacts in LUS images. In order to reduce the number of false and missed detection, our method includes a two-stage validation mechanism, which is performed in both Radon and image domains. We evaluate the performance of the proposed method in comparison to the current state-of-the-art B-line identification method, and show a considerable performance gain with 87\% correctly detected B-lines in LUS images of nine COVID-19 patients.
\end{abstract}


\IEEEpeerreviewmaketitle

\section{Introduction}
\IEEEPARstart{T}{he} outbreak of SARS-CoV-2 at the end of 2019 has rapidly spread to multiple countries on all continents, within the span of just few months. As of the beginning of August 2020, there have been more than 18 million confirmed cases globally with a death toll of more than 700 thousands people, and these numbers are continuously growing. Current approaches to diagnose COVID-19 are based on real-time reverse-transcriptase polymerase chain reaction, which became the gold standard for confirming the infection. However, the sensitivity of this approach is low, and it is mainly useful at the inception of COVID-19.

Some medical imaging modalities, including Computed Tomography (CT) and X-ray, play a major role in confirming positive COVID-19 patients, but use ionising radiation and require patient movement. On the other hand, medical ultrasound is a technology that has advanced tremendously in recent years and is increasingly used for lung problems that previously needed large X-ray or CT scanners. Indeed, lung ultrasound (LUS) can help in assessing the fluid status of patients in intensive care as well as in deciding management strategies for a range of conditions, including COVID-19 patients.

LUS can be conducted rapidly and repeatably at the bedside to help assess COVID-19 patients in intensive care units (ICU), and in emergency settings \cite{thomas2020lung,silva2017combined,bataille2015accuracy}. LUS provides real-time assessment of lung status and its dynamic interactions, which are disrupted in pathological states. In the right clinical context, LUS information can contribute to therapeutic decisions based on more accurate and reproducible data. Additional benefits include a reduced need for CT scans and therefore shorter delays, lower irradiation levels and cost, and above all, improved patient management and prognosis due to innovative LUS quantitative and integrative analytical methods.

The common feature in all clinical conditions, both local to the lungs (e.g. pneumonia) and those manifesting themselves in the lungs (e.g. kidney disease~\cite{anantrasirichai2017line}), is the presence in LUS of a variety of line artefacts. These include so-called A-,B-, and Z-lines, whose detection and quantification is of extremely significant clinical relevance. Bilateral B-lines are commonly present in lung with interstitial oedema. Sub-pleural septal oedema is postulated to provide a bubble-tetrahedron interface, generating a series of very closed spaced reverberations at a distance below the resolution of ultrasound which is interpreted as a confluent vertical echo which does not fade with increasing depth \cite{dietrich2016lung, soldati2016physical, anantrasirichai2017line}. The presence or absence of B-lines in thoracic ultrasonography, as well as their type and quantity, can be used as a marker of COVID-19 disease \cite{soldati2020proposal, buonsenso2020novel,thomas2020lung, peng2020findings,vetrugno2020our}. As reported in \cite{soldati2020proposal} and \cite{peng2020findings}, a thick and irregularly shaped (broken) pleural line, and the existence of focal, multifocal, and/or confluent vertical B-lines are important indicators of the stage of COVID-19, whilst A-lines become visible predominantly during the recovery phase.

A clinical limitation when using LUS is that quantification of line artefacts relies on visual estimation and thus may not accurately reflect generalised fluid overload or the severity of conditions such as interstitial lung disease or severe pneumonia. Moreover, currently the technique is operator-dependent and requires specialist training. Therefore, reliable image processing techniques that improve the visibility of lines and facilitate line detection in speckle images are essential. However, only a few automatic approaches have been reported in the literature \cite{brattain2013automated, moshavegh2016novel, weitzel2015quantitative, anantrasirichai2017line, moshavegh2018automatic}. The method in \cite{brattain2013automated} employs angular features and thresholding (AFT). A B-line is detected in a particular image column if each feature exceeds a predefined threshold. The method in \cite{moshavegh2016novel} uses alternate sequential filtering (ASF). A repeated sequential morphological opening and closing approach is applied to the mask until potential B-lines are separated. In \cite{anantrasirichai2017line}, an inverse problem based method has been proposed for line detection in ultrasound images, referred to as the Pulmonary Ultrasound Image analysis method (PUI) hereafter. It involves $\ell_p$-norm regularisation, which promotes sparsity via small norm orders $p > 0$, and is shown to be suitable for the detection of B-line, as well as of other line artefacts, including Z-lines and A-lines. More recently, deep learning approaches have also been employed in order to detect and localise line artefacts in LUS~\cite{van2019localizing,wang2019quantifying}. The method in \cite{van2019localizing} in particular is able to detect in real-time the presence of B-lines and requires only a small amount of annotated training images.

In this paper, we present a new LUS image analysis system based on a sophisticated non-convex regularisation method for automatic quantification of line artefacts, and validate it on the novel coronavirus (SARS-CoV-2) induced pneumonia.
The proposed non-convex regularisation scheme, which we prefer to call Cauchy proximal splitting (CPS) can be guaranteed to converge following \cite{karakucs2019cauchy1}, whilst benefiting from the advantages offered by a non-convex penalty function. Furthermore, following reconstruction via CPS, a novel automatic line artefacts quantification procedure is proposed, which offers an unsupervised framework for the detection of horizontal (pleural, and either sub-pleural or A-lines), and vertical (B-lines) line artefacts in LUS measurements of several COVID-19 patients.
The performance of the proposed algorithm is compared to the state-of-the-art B-line detection method, PUI \cite{anantrasirichai2017line}. We show B-line quantification performance of 87\% with a gain of around 8\% over PUI \cite{anantrasirichai2017line}. 

The remainder of the paper is organised as follows: the clinical relationship between LUS line artefacts and COVID-19 disease is presented in Section \ref{sec:sec2}. The Cauchy proximal splitting algorithm and the proposed line artefact detection method are presented in detail in Sections \ref{sec:Cauchy} and \ref{sec:LAQMethod}, respectively. Section \ref{sec:data} presents the LUS data sets of COVID-19 patients, whilst experimental results and discussions are presented in Section \ref{sec:results}. Section \ref{sec:conc} concludes the paper with remarks and future work directions.

\section{Line Artefacts and Covid-19}\label{sec:sec2}
\subsection{Line Artefacts in LUS}
Lung ultrasound, is a non-invasive, easy-to-perform, radiation-free, fast, cheap and highly reliable technique, which is currently employed for objective monitoring of pulmonary congestion \cite{sherman2016crackles}. The technique requires ultrasound scanning of the anterior right and left chest, from the second to the fifth intercostal space, in multiple intercostal spaces \cite{jambrik2004usefulness}. The soft tissues of the chest wall and the aerated lung are separated by a pleural line, which is thin, hyperechoic and curvilinear.

Due to the presence in the lung of transudates, the air content decreases and the lung density increases. When this happens, the acoustic difference between the lung and the surrounding tissues is reduced, and the ultrasound beam is reflected multiple times at higher depths. This phenomenon creates discrete vertical hyperechoic reverberation artefacts (B-lines), that arise from the pleural line \cite{gargani2014lung}. The presence of a few scattered B-lines can be normal, and is sometimes encountered in healthy subjects, especially in the lower intercostal spaces. Multiple B-lines are considered sonographic sign of lung interstitial syndrome, and their number increases along with decreasing air content and increase in lung density \cite{trezzi2013lung}. Note that the B-lines are counted as one if they originate from the same point on the pleural line.

\begin{figure}[t]
\centering
\includegraphics[width=\linewidth]{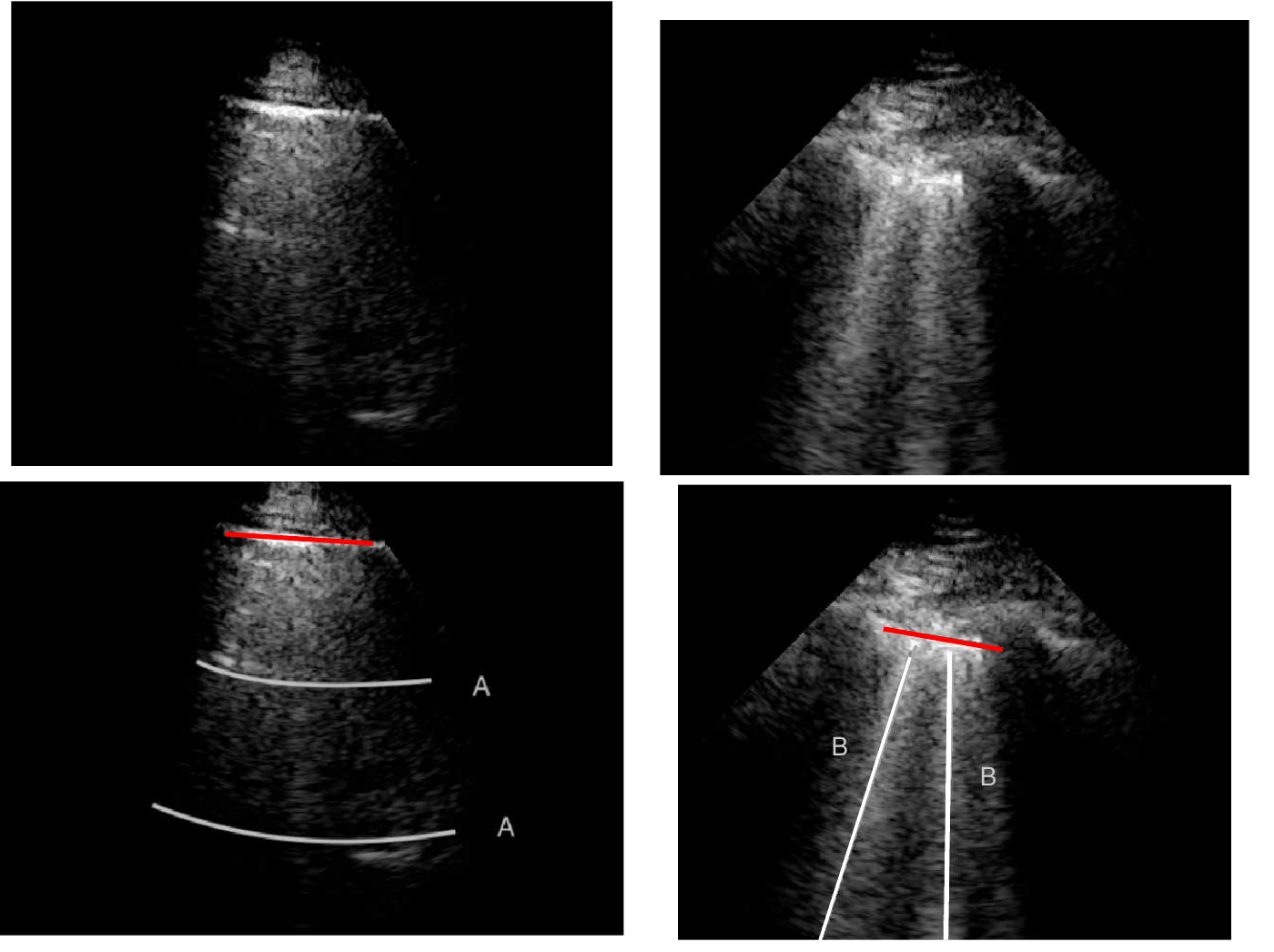}
\caption{Lung ultrasound images
(top row) and visible line artefacts overlaid on them (bottom row). There are two A-lines in the image on the left. There are two B-lines and a single A-line in the image on the right.}
\label{fig:exLines}
\end{figure}

A-line artefacts are repetitive horizontal echoic lines with equidistant intervals, which are also equal to the distance between skin and pleural line \cite{lichtenstein2009lines}. The A-lines indicate sub-pleural air, which completely reflects the ultrasound beam. The length of an A-line can be roughly the same as the pleural line, but it can also be shorter, or even not visible because of sound beam attenuation through the lung medium. An example depicting the A- and B-line artefacts in LUS is presented in Fig. \ref{fig:exLines}.

\subsection{Clinical Significance for  COVID-19 Patients}
As previously mentioned, line artefacts are highly likely to be observed during SARS-CoV-2 infection in LUS images. Detection of such lines helps to rapidly and repeatably assess COVID-19 patients at the bedside, in ICU or in emergency services. Due to its efficacy in diagnosing adult respiratory distress syndrome (ARDS) and pneumonia, LUS has been extensively used to evaluate COVID-19 patients \cite{peng2020findings}.
There are several important clinical findings associated to LUS line artefacts in COVID-19 patients, such as:
\begin{itemize}
    \item A thick and irregularly shaped pleural line,
    \item Focal, multifocal and confluent B-lines,
    \item A-lines in the recovery phase.
\end{itemize}

Specifically, depending on the stage of the disease and severity of the lung injury, line artefacts with different structures are visible in LUS for COVID-19 patients. In early stages and mild cases, focal B-lines are common. In progressive stage and for critically ill patients, one main characteristic is the alveolar interstitial syndrome. Patients with pulmonary fibrosis are most likely to have a thickened pleural line and uneven B-line artefacts. A-lines are generally observed in patients during the recovery stage \cite{peng2020findings}.

\section{Non-convex regularisation for Line Artefacts Detection}\label{sec:Cauchy}
\subsection{Line Artefacts and the LUS Image Formation Model}
The LUS image formation model involving line artefacts can be expressed as \cite{anantrasirichai2017line}
\begin{align}\label{equ:wake}
    Y = \mathcal{C}X + N
\end{align}
where $Y$ is the ultrasound image, $N$ is additive white Gaussian noise (AWGN) and $\mathcal{C} = \mathcal{R}^{-1}$ is the inverse Radon transform operator. $X(r, \theta)$ refers to lines as a distance $r$ from the centre of $Y$, and an orientation $\theta$ from the horizontal axis of $Y$. 

In image processing applications, computing the integral of the intensities of image $Y(i, j)$ (where $i$ and $j$ refer to the pixel locations with respect to rows, and columns, respectively) over the hyperplane perpendicular to $\theta$, corresponds to the Radon transform $X(r, \theta)$ of the given image $Y$. It can also be defined as a projection of the image along the angles, $\theta$. Hence, for a given image $Y$, the general form of the Radon transform ($X=\mathcal{R}Y=\mathcal{C}^TY$) is
\begin{align}\label{equ:R}
  X(r, \theta) = \int_{\mathds{R}^2} Y(i, j) \delta(r - i\cos\theta - j\sin\theta) didj
\end{align}
where $\delta(\cdot)$ is the Dirac-delta function.

The inverse Radon transform ($Y=\mathcal{C}X$) of the projected image $X$ can be obtained from the filtered back-projection \cite{avinash1988principles} algorithm as
\begin{align}\label{equ:C}
  g(r, \theta) &= \mathcal{F}^{-1} \left[ |v| \mathcal{F} \left[ X(r, \theta) \right] \right]\\
  Y(i, j) &= \int_{0}^{\pi} \int_{\mathds{R}} g(r, \theta) \delta(r - i\cos\theta - j\sin\theta) drd\theta
\end{align}
where $v$ is the radius in Fourier transform, and $\mathcal{F}\left[\cdot\right]$ and $\mathcal{F}^{-1}\left[\cdot\right]$ refer to forward and inverse Fourier transforms, respectively. In this work, discrete operators $\mathcal{R}$ and $\mathcal{C}$ are used as described in \cite{kelley1993fast}.

\subsection{Non-Convex Regularisation with a Cauchy Penalty}
In this section, we propose the use of the Cauchy distribution in the form of a non-convex penalty function for the purpose of inverting the image formation model given in \eqref{equ:wake}. The Cauchy distribution is one of the special members of the $\alpha$-stable distribution family which is known to be heavy-tailed and to promote sparsity in various applications \cite{wan2011segmentation,yang2020detection,karakucs2020solving}. It thus fits our problem of detecting a few lines in an image well. Contrary to the general $\alpha$-stable family, it has a closed-form probability density function, which is given by \cite{karakucs2019cauchy1}
\begin{align} \label{equ:Cauchy}
    p(x) \propto \frac{\gamma}{\gamma^2+x^2}
\end{align}	
where $\gamma$ is the scale (or the dispersion) parameter, which controls the spread of the distribution. 

Considering the model in (\ref{equ:wake}) with the Cauchy prior given in (\ref{equ:Cauchy}), based on a Maximum A Posteriori estimation formulation, the proposed method reconstructs $X$ via \cite{karakucs2019cauchy1}
\begin{align}\label{equ:miniCauchy}
    \hat{X}_{\text{Cauchy}} = \arg\min_X \frac{\|Y - \mathcal{C}X\|_2^2}{2\sigma^2} - \sum_{i,j} \log\left(\frac{\gamma}{\gamma^2+X_{ij}^2}\right).
\end{align}
The function to be minimised is composed of two terms. The first is an $\ell_2$-norm data fidelity term resulting from the AWGN assumption. The second is the non-convex Cauchy-based penalty function originally proposed in \cite{karakucs2019cauchy1}:
\begin{align}\label{equ:neglogCuachy}
\psi(x) = -\log\left(\dfrac{\gamma}{\gamma^2+x^2}\right).
\end{align} 

In order to solve the minimisation problem in (\ref{equ:miniCauchy}) by using proximal algorithms such as the forward-backward (FB) or the alternating direction method of multipliers (ADMM), \textit{the proximal operator} of the Cauchy based penalty function should be defined.
In the related recent publication \cite{karakucs2019cauchy1}, we proposed a closed-form expression for the proximal operator of the Cauchy based penalty function (\ref{equ:neglogCuachy})
\begin{align}\label{equ:proxCauchy}
    prox_{Cauchy}^{\mu}(x) = \arg\min_u \left\{\frac{\|x - u \|_2^2}{2\mu} - \log\left(\frac{\gamma}{\gamma^2+u^2}\right) \right\}.
\end{align}

The solution to this minimisation problem can be obtained by taking the first derivative of (\ref{equ:proxCauchy}) in terms of $u$ and setting it to zero. Hence we have 
\begin{align}\label{equ:proxCauchy2}
    u^3-xu^2+(\gamma^2+2\mu)u-x\gamma^2 = 0.
\end{align}

The solution to the cubic function given in (\ref{equ:proxCauchy2}) can be obtained through Cardano’s method as \cite{karakucs2019cauchy1}
\begin{align}
    p &\gets \gamma^2 + 2\mu - \frac{x^2}{3},\\
q &\gets x\gamma^2 + \frac{2x^3}{27} - \frac{x}{3}\left(\gamma^2 + 2\mu\right),\\
s &\gets \sqrt[3]{q/2 + \sqrt{p^3/27 + q^2/4}},\\
t &\gets \sqrt[3]{q/2 - \sqrt{p^3/27 + q^2/4}},\\
\label{equ:cardano}z &\gets \frac{x}{3} + s + t.
\end{align}
where $z$ is the solution to $prox_{Cauchy}^{\mu}(x)$.

\begin{figure*}[t]
\centering
\includegraphics[width=\linewidth]{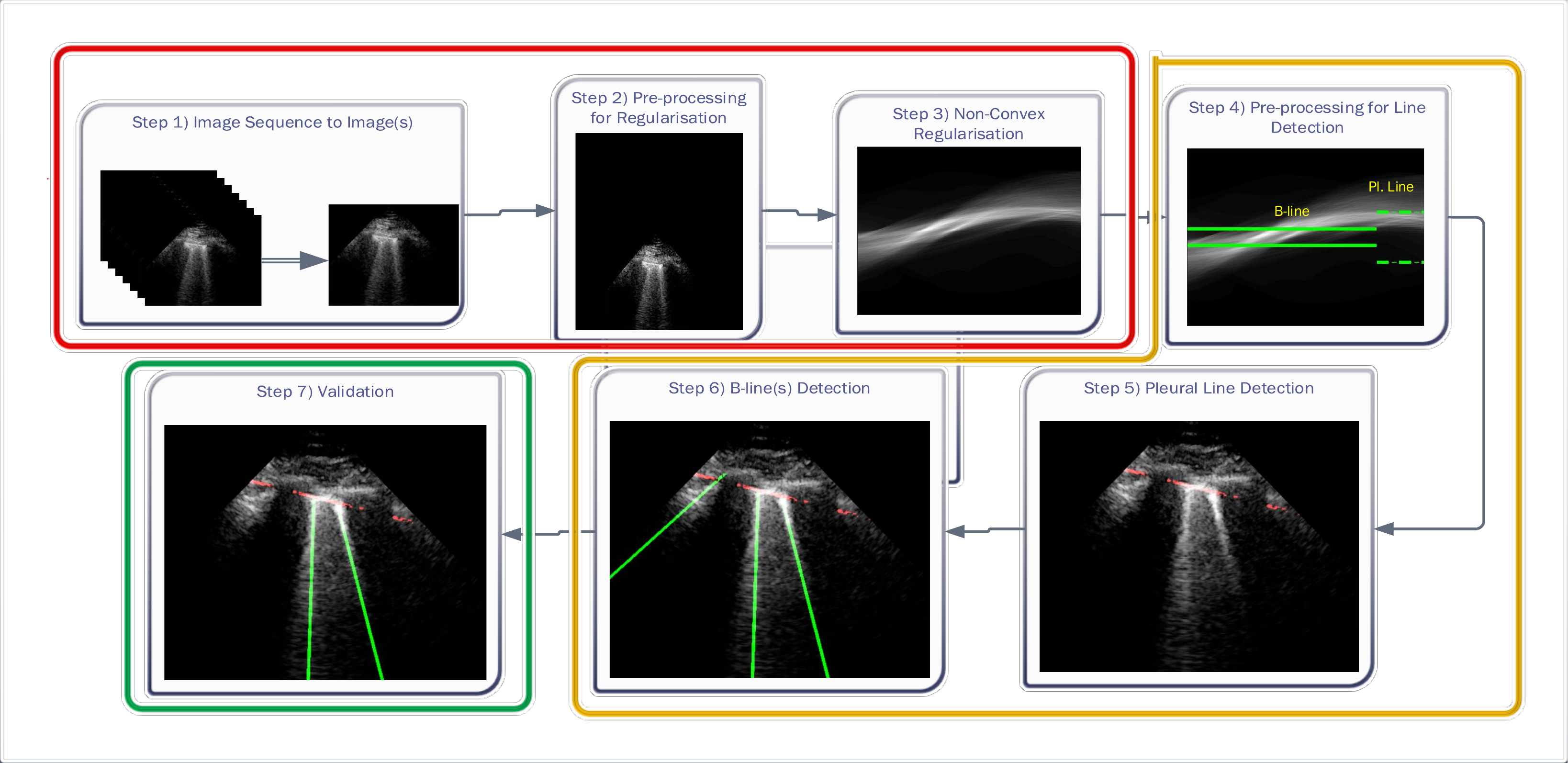}
\caption{Schematic view of the proposed line artefact quantification algorithm.}
\label{fig:det}
\end{figure*}

\subsection{Cauchy Proximal Splitting}
The use of a proximal operator corresponding to the proposed penalty function would enable the use of a proximal splitting algorithm to solve the optimisation problem in (\ref{equ:miniCauchy}).
In particular, an optimisation problem of the form
\begin{align}\label{equ:FB1}
    \arg\min_x (f_1 + f_2)(x)
\end{align}
can be solved via the FB algorithm. From the definition \cite{combettes2011proximal}, provided $f_2:\mathbb{R}^N \rightarrow \mathbb{R}$ is $L$-Lipchitz differentiable with Lipchitz constant $L$ and $f_1:\mathbb{R}^N \rightarrow \mathbb{R}$, then (\ref{equ:FB1}) can be solved iteratively as
\begin{align}\label{equ:FB2}
    x^{(n+1)} = prox_{f_1}^{\mu} \left( x^{(n)} - \mu\bigtriangledown f_2(x^{(n)})  \right),
\end{align}
where the step size $\mu$ is set within the interval $\left(0, \frac{2}{L} \right)$. In our case, the function $f_2$ is the data fidelity term and takes the form of $\frac{\|y - \mathcal{C}x\|_2^2}{2\sigma^2}$ from (\ref{equ:miniCauchy}) whilst the function $f_1$ corresponds to the Cauchy based penalty function $\psi$, proven in~\cite{karakucs2019cauchy1} to be twice continuously differentiable. 

Observing (\ref{equ:miniCauchy}), it can be easily deduced that since the penalty function $\psi$ is non-convex, the overall cost function is also non-convex in general. Hence, in order to avoid local minimum point estimates, one should ensure convergence of the proximal splitting algorithm employed. To this effect, we have formulated the following theorem in \cite{karakucs2019cauchy1}, which we recall here for completeness.

\begin{thm}
 \label{thm:theorem2}
Let the twice continuously differentiable and non-convex regularisation function $\psi$ be the function $f_1$ and the $L$-Lipchitz differentiable data fidelity term $\frac{\|y - \mathcal{C}x\|_2^2}{2\sigma^2}$ be the function $f_2$. The iterative FB sub-solution to the optimisation problem in (\ref{equ:miniCauchy}) is
\begin{align}\label{equ:thm2}
    x^{(n+1)} = prox_{Cauchy}^{\mu} \left( x^{(n)} - \frac{\mu\mathcal{C}^T(\mathcal{C}x^{(n)} - y)}{\sigma^2}  \right)
\end{align}
where $\bigtriangledown f_2(x^{(n)}) = \frac{\mathcal{C}^T(\mathcal{C}x^{(n)} - y)}{\sigma^2}$. If the condition
\begin{align}\label{equ:Proofthm2}
    \gamma \geq \frac{\sqrt{\mu}}{2}
\end{align}
holds, then the sub-solution of the FB algorithm is strictly convex, and the FB iteration in (\ref{equ:thm2}) converges to the global minimum.
\end{thm}

For the proof of the theorem, we refer the reader to \cite{karakucs2019cauchy1}.
In order to comply with the condition imposed by the theorem, two approaches are possible: (i) the step size $\mu$ can be set following estimation of $\gamma$ directly from the observations, (ii) the scale parameter $\gamma$ can be set, for cases when the Lipchitz constant $L$ is computed or if estimating $\gamma$ requires computationally expensive calculations. In this paper, we follow the second option, i.e. (calculate $L$) $\rightarrow$ (set $\mu\in\left(0, \frac{2}{L}\right)$) $\rightarrow$ (set $\gamma\geq \frac{\sqrt{\mu}}{2}$).

Based on Theorem \ref{thm:theorem2}, Algorithm \ref{alg:FB} presents the proposed CPS algorithm for solving (\ref{equ:miniCauchy}).

\begin{algorithm}[ht!]
\caption{The CPS Algorithm}\label{alg:FB}
\setstretch{1.2}
\begin{algorithmic}[1]
\State \textbf{Input:} $\text{Input image, }Y \text{ and } MaxIter$
\State \textbf{Input:} $\mu\in\left(0, \frac{2}{L}\right) \text{ and } \gamma\geq\frac{\sqrt{\mu}}{2}$
\State \textbf{Output:} $X$
\State \textbf{Set:} $i=0 \text{ and } X^{(0)}$
  \Do
    \State $u^{(i)} = X^{(i)} - \mu \mathcal{C}^T(\mathcal{C}X^{(i)} - Y)$
    \State $X^{(i+1)} = prox_{Cauchy}^{\mu}(u^{(i)})\text{ from } (\ref{equ:cardano})$
    \State $i++$
  \doWhile{$\dfrac{\|X^{(i)} - X^{(i-1)}\|}{\|X^{(i-1)}\|} > \varepsilon \text{ or } i<MaxIter$}
\end{algorithmic}
\end{algorithm} 

\section{Line Artefact Quantification}\label{sec:LAQMethod}
The proposed line artefact detection method in this paper comprises of three main stages with seven different steps within. Figure \ref{fig:det} depicts the proposed algorithm, in which coloured boxes represent states: 1) Pre-processing and non-convex regularisation (red box), 2) Detecting Line Artefacts (orange box), and 3) Validation (green box). We discuss each stage in detail in the sequel.

\subsection{Pre-processing and Non-convex Regularisation}
The first stage of the detection method is the pre-processing and non-convex regularisation, which covers the first three steps.
\subsubsection*{Step 1) Image sequence to Image(s)}
The given LUS image sequence is decomposed into image frames to be processed. If the image sequence includes some informing text or other scanner related information, all these are removed, so that only ultrasound can be processed by the following steps.
\subsubsection*{Step 2) Pre-processing for regularisation}
\begin{figure}[htbp]
\centering
\subfigure[]{\includegraphics[width=.21\linewidth]{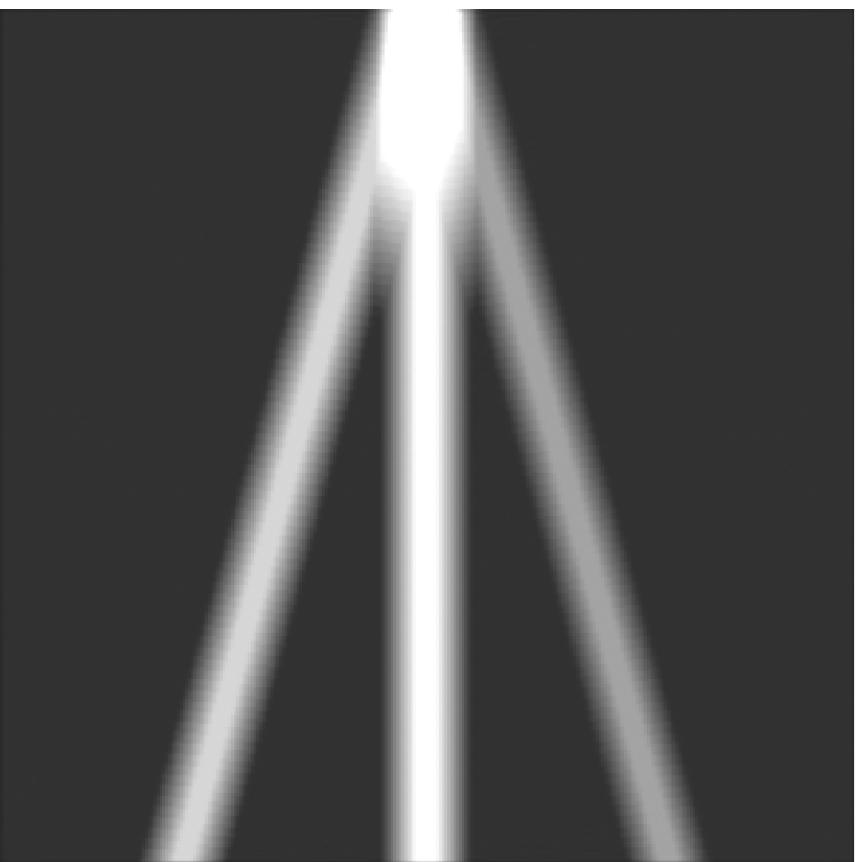}}
\subfigure[]{\includegraphics[width=.42\linewidth]{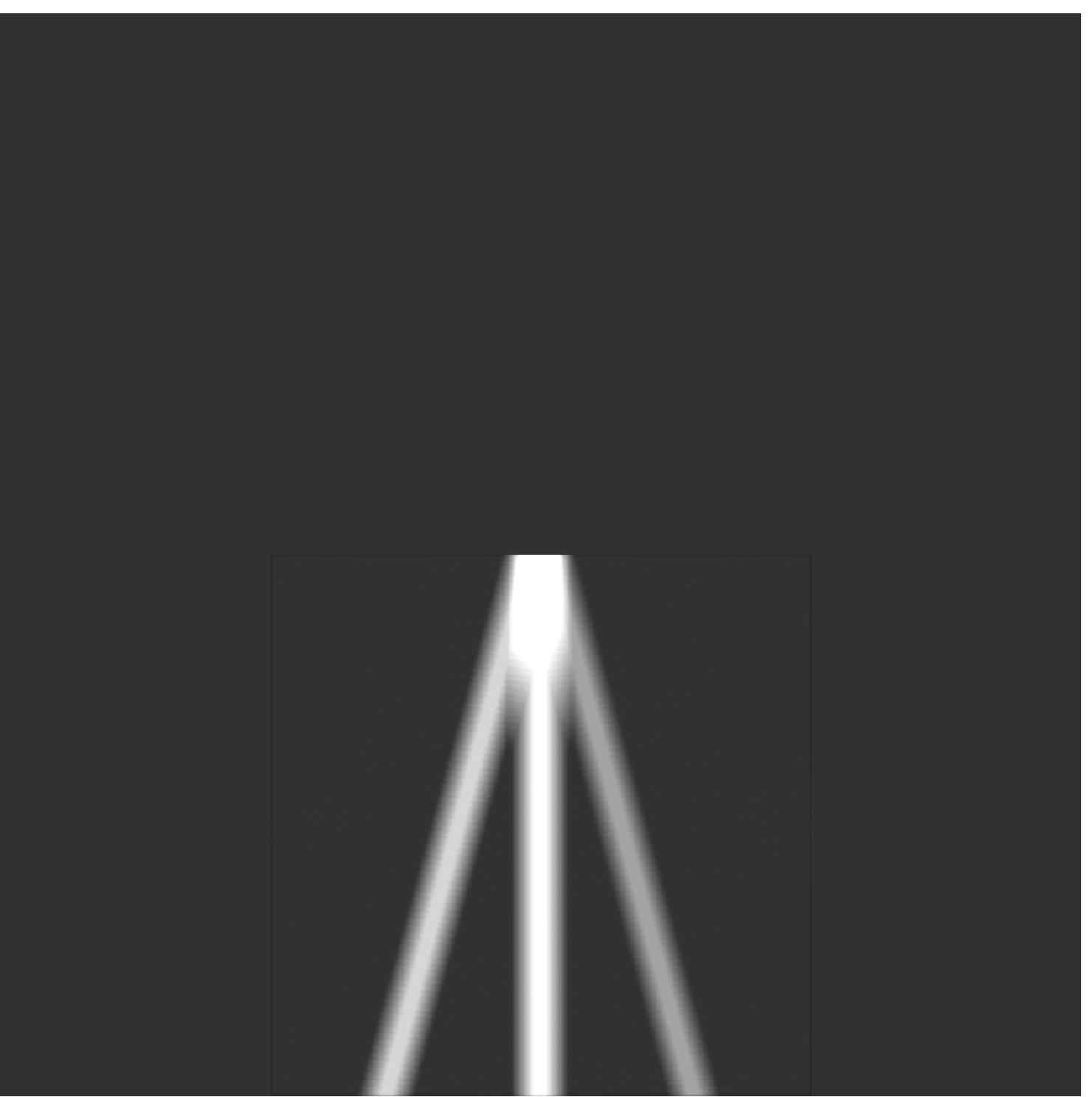}}\\
\subfigure[]{\includegraphics[width=.32\linewidth]{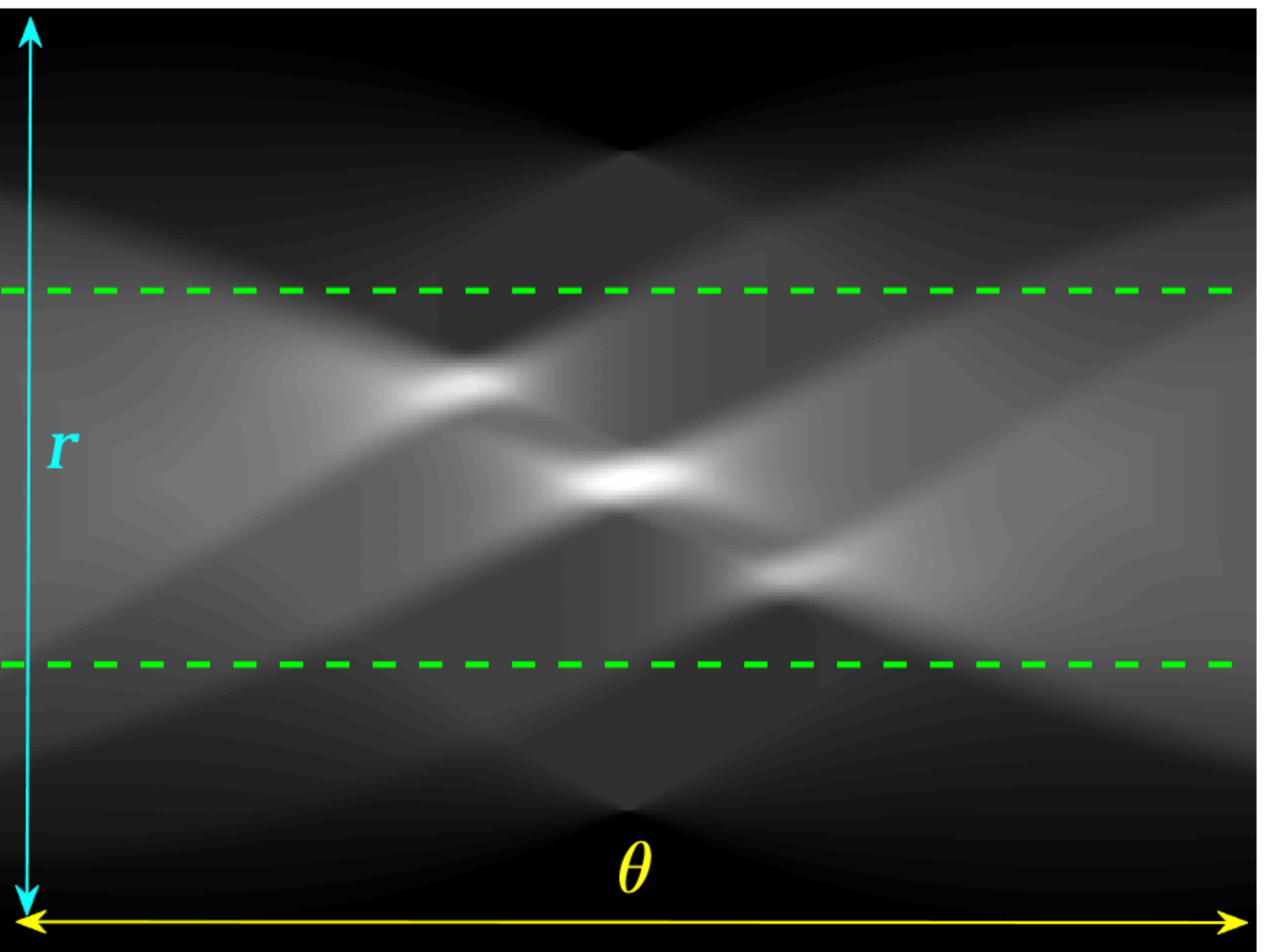}}
\subfigure[]{\includegraphics[width=.32\linewidth]{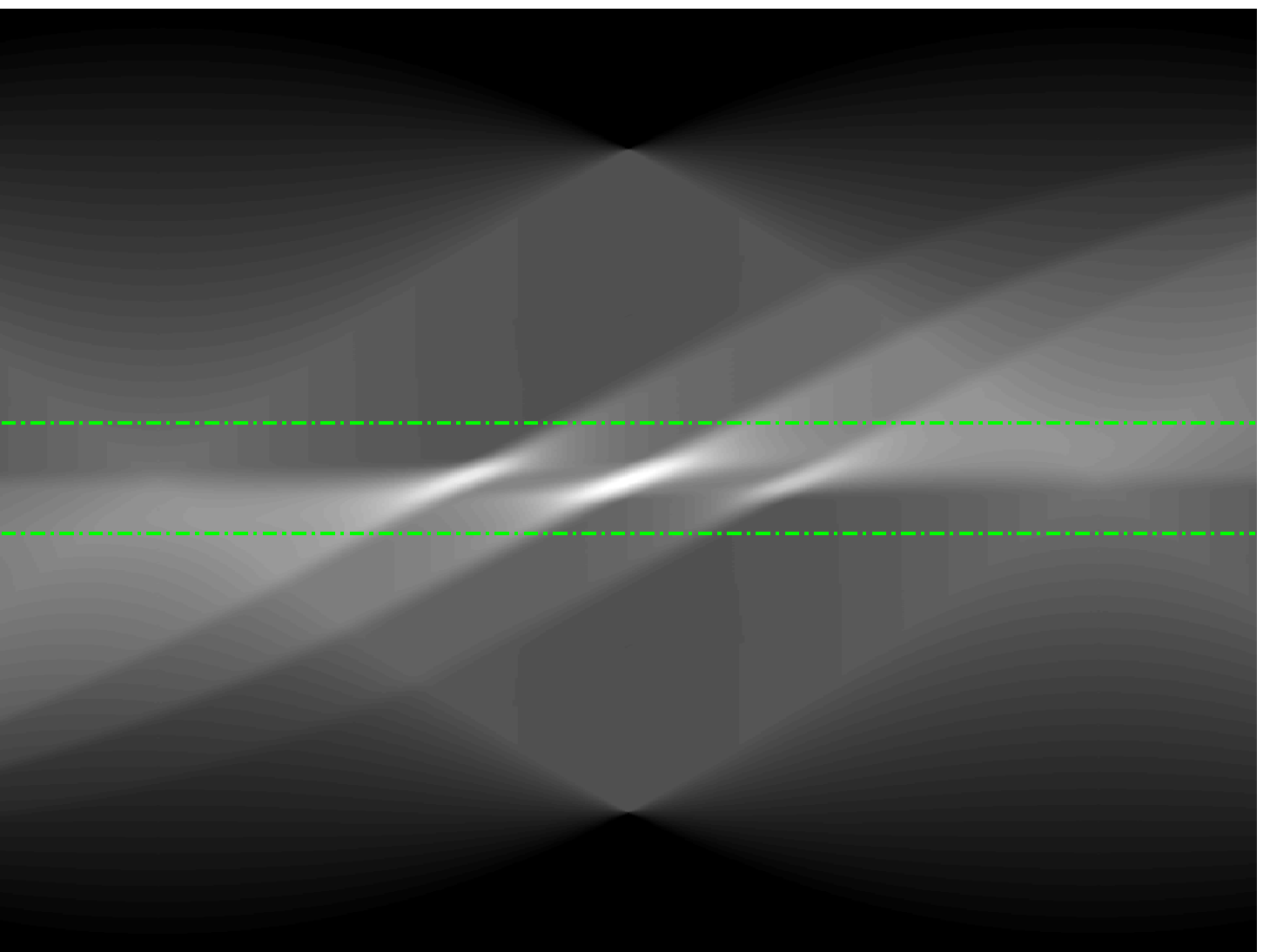}}
\caption{A visual representation of Step 2 in the proposed line artefact detection algorithm. (a) Simulated US image. (b) Simulated "probe centred" image. (c) Radon transform of (a). (d) Radon transform of (b). Dotted lines in (c) and (d) refer to search area limitation performed in Step 4. In (c), Radon space components $\theta$ and $r$ are also shown.}
\label{fig:step2}
\end{figure}

A pre-processing step is then applied to each image frame before employing the non-convex regularisation algorithm. Specifically, the pre-processing procedure includes a transformation in the image domain, which creates a "probe centred" image. An example of this procedure is presented in Figure \ref{fig:step2}. An image frame extracted from an image sequence (set of frames, or video) is first considered. Then, the probe centre of the considered image frame is located in the centre of a new template image. In Figure \ref{fig:step2}-(b) the probe centred image corresponding to the image frame in (a) is shown. The reason behind this operation is coming from a process we apply for detecting candidate B-lines in Radon transform domain. Note that a similar idea has been proven to be successful in an application of ship wake detection of synthetic aperture radar imaging of the sea surface in \cite{karakucs2019ship2,graziano1}. An example for the effect of the procedure in this step in Radon space is also shown in Figures \ref{fig:step2}-(c) and (d). As it can be clearly seen from (d), the corresponding peak points in Radon domain (corresponding to candidate B-lines in image domain) for probe centred image are located in the middle of the vertical Radon domain axis, whilst they cover a larger area in (c). This helps to limit the search area in Radon domain to a smaller useful range, which increases the detection accuracy and reduces the number of false detection because of noisy peak points in Radon domain. Although this procedure doubles the input image size, which causes a higher computational load, this is counterbalanced by a significant increase in detection performance.

\subsubsection*{Step 3) Non-Convex regularisation} The probe centred image obtained in Step 2 is subsequently fed as the input LUS image $Y$ of the model in (\ref{equ:wake}). Thence, the proposed non-convex regularisation with the Cauchy based penalty function in (\ref{equ:miniCauchy}) is performed to promote sparsity of the linear structures in the Radon domain. The output of the 3$^{rd}$ step in the 1$^{st}$ stage of the algorithm consists in the reconstructed Radon space information of the linear structures, $\hat{X}_{Cauchy}$.

\subsection{Detecting Line Artefacts}
The second stage in the line artefact detection algorithm is the detection of all horizontal and vertical lines in Radon domain, which includes the steps from 4 to 6 in Figure \ref{fig:det}.

\subsubsection*{Step 4) Pre-processing for Line Detection in Radon Domain} Following on from Step 2, Radon space information reconstructed via the CPS algorithm is then used to detect candidate horizontal (pleural, sub-pleural or A) and vertical (B) lines. The pre-processing step consists of limiting search areas in Radon domain for both horizontal and vertical linear structures. Note that all the detected peak points in this stage are first considered as "candidate lines", since a thresholding/validation procedure is performed in the following steps. Furthermore, all detected B-lines become final  only after the validation process performed in Step 7.

Deciding upper and lower borders for search areas has a crucial importance in the detection process. Selecting a large value increases the searching area, which simultaneously increases not only the number of possible candidate linear structures, but also the possibility of false detection of noisy peaks. Conversely, selecting small values may cause lines to fall outside of the search area, leading to an increase of the number of missed detections. Hence, we use borders of the search areas defined as:
\begin{align}
    \text{Horizontal lines search area} \rightarrow& \quad r \leq \left|\frac{M}{4}\right|,\\
    \text{B-lines search area} \rightarrow& \quad r \leq \left|\frac{M}{16}\right|,
\end{align}
where $M$ is the size of the probe centred LUS image $Y$ ($M\times M$). These borders were experimentally set to lead to the best detection results. Since the vertical lines generally start from the probe centre, the search area for vertical lines become naturally smaller than the horizontal ones.

\subsubsection*{Step 5) Pleural (Horizontal) Line Detection} As mentioned above, the second stage of the line artefact quantification method proposed in this study addresses the identification of several linear structures, corresponding to both horizontal lines (pleural, and either sub-pleural or A-lines) and vertical B-lines. The method first detects the pleural line to locate the lung space where line artefacts occur. Since the pleural line is generally horizontal, we limit the searching angle $\theta_p$ within the range $[60^{\circ}\text{, }90^{\circ}]$. Then, the horizontal search area given above is scanned for detecting the local peaks of the reconstructed Radon information. All detected local peak points are candidate horizontal lines. Firstly, we select the maximum intensity local peak, which is the closest to the Radon domain vertical axis centre. The corresponding peak point is set as the detected pleural line. Following the pleural detection, the proposed method then searches $m$ more horizontal lines without classifying them yet as sub-pleural or A-lines. The user defined value $m$ refers to the number of horizontal lines (except the pleural line) to be searched. The same detection procedure is then employed for detecting $m$ other horizontal lines.   

\subsubsection*{Step 6) B-line Detection} 
\begin{figure}[t]
\centering
\includegraphics[width=0.99\linewidth]{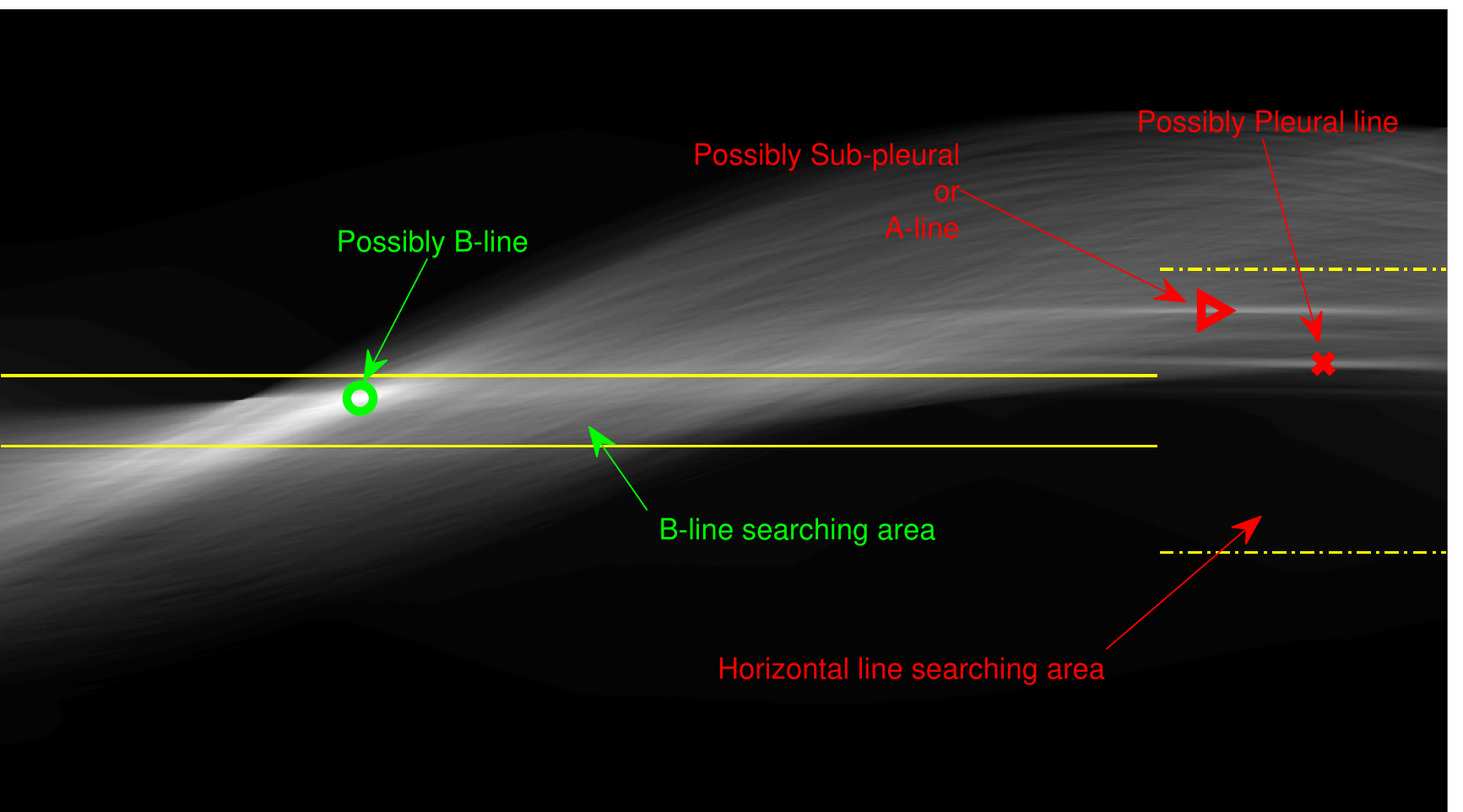}
\caption{An example of Radon domain processing for Stage-2 of the proposed line artefact method.}
\label{fig:stage2}
\end{figure}

Following the detection of horizontal lines, the procedure for B-line identification starts with detecting the vertical lines in the image by limiting  the search angle $\theta_p$ within the range $[-60^{\circ}\text{, }60^{\circ}]$. This is done by first searching for local peaks and adopting them as candidate B-lines. At this point, there is a high number of candidate B-lines, most of which are just local peaks not corresponding to true B-lines. Hence, in order to eliminate potential false detections, a thresholding operation is needed. Similar to \cite{anantrasirichai2017line}, the proposed method uses the following threshold procedure:
\begin{align}\label{equ:TH}
    \begin{tabular}{l}
    \textbf{\textit{if}} $X(r_v, \theta_v) > \frac{h_{lung}}{2}$ \\
    \hspace{0.5cm} Keep candidate point $(r_v, \theta_v)$, \\
    \textbf{\textit{else}} \\
    \hspace{0.5cm} Eliminate candidate point $(r_v, \theta_v)$. \\
    \textbf{\textit{end}} \\
    \end{tabular}
\end{align}
where $r_v$ and $\theta_v$ are the detected peak coordinates in Radon space, and $h_{lung}$ refers to the distance between the pleural line and the bottom of the image. Although the definition of B-lines assumes that they extend to the bottom of the ultrasound image, in some cases this might not be true because of amplitude attenuation, which is not compensated perfectly. Hence, selecting the threshold as $h_{lung}/2$ has proven to be adequate in \cite{anantrasirichai2017line}.
Following the thresholding operation, we reduce the number of candidate B-lines and the final detected number of B-lines is decided after the validation process in Step 7. An example of Radon domain processing for steps 4, 5 and 6 in Stage-2 of the proposed line artefact method is shown in Figure \ref{fig:stage2}.

\subsection{Validation}
The third and last stage of the line artefact quantification method is the validation, which consists of Step 7 in Fig. \ref{fig:det}. 
\subsubsection*{Step 7) Validation} Even though we use a thresholding operation to eliminate erroneous peak detections in Radon space, in order to reduce the number of false detection of B-lines, we perform another validation step, this time in the image domain. Specifically, the validation of the candidate B-lines starts by calculating a measure index $F$, which is given by:
\begin{align}
    F = \dfrac{\bar{I}_{line}}{\bar{I}_{LUS}} - 1,
\end{align}
where $\bar{I}_{line}$ is the average intensity over the detected candidate B-line, and $\bar{I}_{LUS}$ is the average image intensity. The $F$ index quantifies linear structures in the image domain, and results in higher values for brighter linear structures and lower values for less visible ones, and possibly false detections. Therefore, deciding a margin will help to reduce the possibility of false confirmations. 
Hence, candidate B-lines which do not follow:
\begin{align}\label{equ:index}
F > F_{val},
\end{align}
are discarded, whereas the remaining B-lines are validated as the final detection results. To this end, please note that setting a high $F_{val}$ value might lead to discarding correct linear structures (missed detection), whilst a lower $F_{val}$ value might increase the number of false detections. On the other hand, since every single LUS image has different levels of intensity, they have different $\bar{I}_{LUS}$ values. Therefore, in order to create a fair margin, we should take into account the average image intensity in calculating the decision criterion. Hence, the corresponding margin is subsequently assumed to be between 25\% and 50\% after a trial-error procedure, according to
\begin{align}
    F_{val} = \min\bigg\{0.5, \max \bigg\{0.25, \dfrac{3}{2}\bar{I}_{LUS} \bigg\} \bigg\}.
\end{align}

\begin{table*}[htbp]
  \centering
  \caption{Clinical characteristics of COVID-19 patients.}
  \begin{tabularx}{0.96\linewidth}{@{} LCCCCCCCC @{}}
  \toprule
    Patient No & Sex   & Age (years) & BMI (kg/m2) & COVID-19 first symptoms & ICU admission & LUS acquisition & Mechanical
Ventilation & PaO2/FiO2
(mmHg) \\ \toprule
    1     & M     & 68    & 28    & 15/03/2020 & 19/03/2020 & 06/04/2020 & Yes   & 42 \\
    2     & M     & 51    & 26.9  & 18/03/2020 & 25/03/2020 & 06/04/2020 & Yes   & 340 \\
    3     & M     & 77    & 25.3  & 16/03/2020 & 18/03/2020 & 06/04/2020 & Yes   & 207 \\
    4     & M     & 68    & 24.9  & 26/03/2020 & 01/04/2020 & 06/04/2020 & Yes   & 87 \\
    5     & F     & 58    & 34.5  & 03/04/2020 & 08/04/2020 & 08/04/2020 & No    & 161 \\
    6     & F     & 67    & 36.5  & 03/04/2020 & 08/04/2020 & 08/04/2020 & Yes   & 184 \\
    7     & M     & 61    & 23.4  & 25/03/2020 & 09/04/2020 & 09/04/2020 & Yes   & 85 \\
    8     & M     & 54    & 32.1  & 26/03/2020 & 07/04/2020 & 09/04/2020 & Yes   & 230 \\
    9     & F     & 79    & 40.8  & 08/04/2020 & 16/04/2020 & 16/04/2020 & Yes   & 113 \\
    \bottomrule
    \end{tabularx}
  \label{tab:patients}%
\end{table*}%

\begin{figure}[t]
\centering
\includegraphics[width=0.99\linewidth]{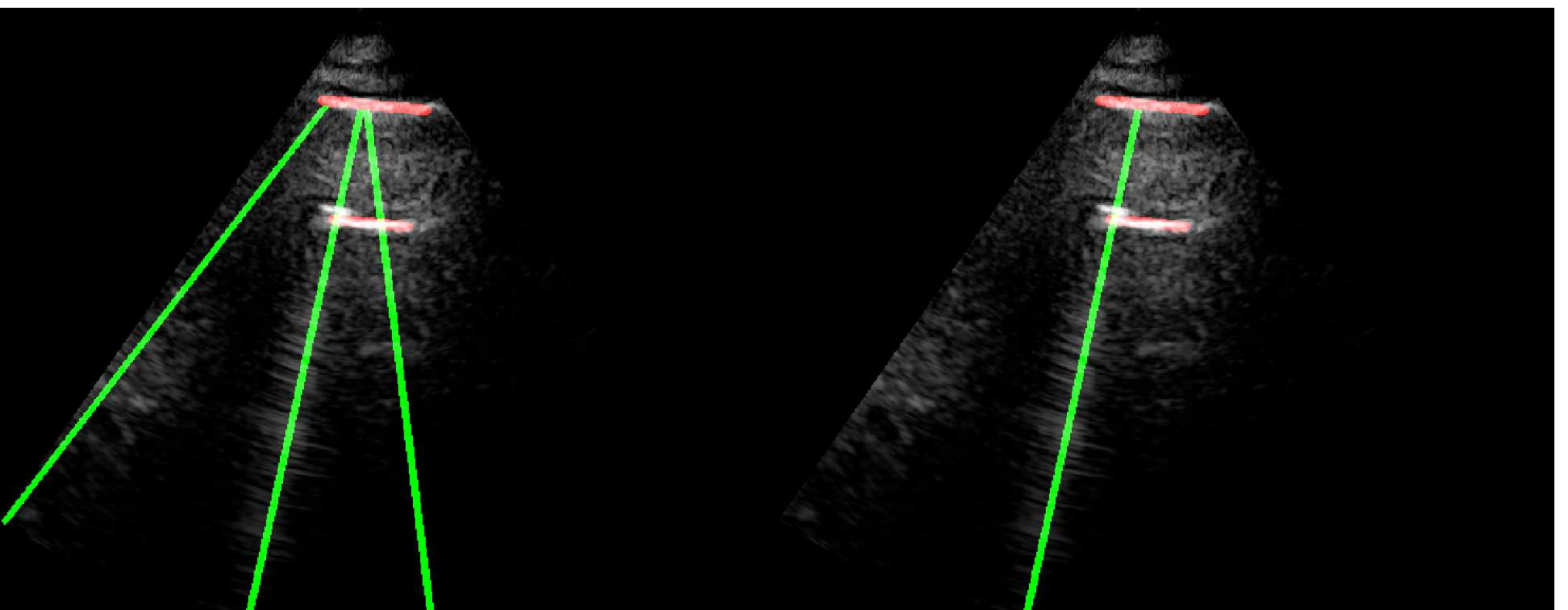}
\caption{An example demonstrating the validation procedure in Step 7. Figure on the left refers to the detection result without validation, whilst the right one is with the validation. The $F$ index values for the detected B-lines from left to right are -0.235, 0.677, and 0.198, respectively. The corresponding validation criterion $F_{val}$ is 0.25 for this example.}
\label{fig:step7}
\end{figure}

An example demonstration of the validation step is depicted in Fig.~\ref{fig:step7}. From the figure, it can be clearly seen that even though a thresholding operation is performed in Radon space in step 6 (Eq. (\ref{equ:TH})), there are still three detected candidate B-lines, two of which are clearly false detections. However, applying the validation operation in this step helps to discard these two false positives as illustrated in the right hand side sub-figure.

\section{Data Set}\label{sec:data}
The datasets employed in this study correspond to nine COVID-19 patients, with clinical details provided in Table \ref{tab:patients}. In particular, we examined 6 male and 3 female patients aged between 51 and 79 years old. All patients were admitted to ICU and 8 of them required mechanical ventilation support.

All patients underwent LUS assessment by investigators who did not take part in their clinical management. Investigators used standardised criteria and followed a pattern analysis. Lung ultrasound was performed with a Philips CX-50 general imaging ultrasound machine (Koninklijke Philips, Amsterdam, Netherlands) and a 1- to 5-MHz bandwidth convex probe. The exploration depth was set at $13$ cm. All patients were investigated in the semirecumbent position.

The anterior chest wall was delineated from the clavicles to the diaphragm and from the sternum to the anterior axillary lines. The lateral chest was delineated from the axillary zone to the diaphragm and from the anterior to the posterior axillary line. Each chest wall was divided into three lung regions. The pleural line was defined as a horizontal hyperechoic line visible 0.5 cm below the rib line. A normal pattern was defined as the presence, in every lung region, of a lung sliding with A lines (A profile). Losses of aeration were described as B-profile. Alveolar consolidation was defined as the presence of poorly defined, wedge-shaped hypoechoic tissue structures (C profile). Each of the 12 lung regions examined per patient was classified in one of these profiles to define specific quadrants.  

\section{Results and Discussions}\label{sec:results}
We consider the experimental part of this paper from two separate angles. Firstly, we evaluated the proposed line artefacts quantification method on LUS images of several COVID-19 patients, described in details in Section \ref{sec:data}. We also compared our method to the state-of-the-art B-line identification method PUI \cite{anantrasirichai2017line}. Secondly, we highlight the image sequence processing capability of the proposed method for LUS measurements corresponding to five patients.

The first example demonstrates the use of the non-convex Cauchy based penalty function and the proposed line artefacts quantification method for 100 different LUS images of nine COVID-19 patients. The B-line detection performance was measured using several metrics regrouped in Table \ref{tab:perfMetrics}.
\begin{table}[htbp]
  \centering
  \caption{Descriptions of performance comparison metrics}
    \begin{tabularx}{0.96\linewidth}{@{} LL @{}}
    \hline
    Expression & Description \\
    \hline
    True Positive (TP) & Correct confirmation of B-lines. \\
    True Negative (TN) & Correct discard of invisible lines \\
    False Positive (FP) & False detection\\
    False Negative (FN) & Missed detection  \\
    Recall & TP/(TP+FN) \\
    Precision & TP/(TP+FP) \\
    Specificity & TN/(TN + FP) \\
    True Positive Rate (TPR) & Equivalent to Recall\\
    False Positive Rate (FPR) & (1 - Specificity)\\
    \% Detection Accuracy & 100(TP+TN)/(TP+FP+TN+FN) \\
    \% Missed Detection & 100(FN)/(TP+FP+TN+FN) \\
    \% False Detection & 100(FP)/(TP+FP+TN+FN) \\
    $F_{\beta}$ & (1 + $\beta^2$)Precision$\times$Recall/($\beta^2$Precision + Recall)\\
    Positive Likelihood & Recall/(1-Specificity)\\
    Ratio (LR+) &\\
    \hline
    \end{tabularx}%
  \label{tab:perfMetrics}%
\end{table}%

\begin{table}[ht]
  \centering
  \caption{Performance metrics for B-line quantification}
    \begin{tabularx}{0.96\linewidth}{@{} LCC @{}}
    \toprule
     Performance     & \multicolumn{1}{c}{The Proposed} & \multicolumn{1}{c}{PUI} \\
     Metric     & \multicolumn{1}{c}{Method} & \multicolumn{1}{c}{\cite{anantrasirichai2017line}} \\
          \toprule
    \% Detection Accuracy & 87.349\% & 78.916\% \\
    \% Missed Detection & 5.422\% & 13.855\% \\
    \% False Detection & 7.229\% & 7.229\% \\
    Specificity & 7.692\% & 14.286\% \\
    Recall & 94.118\% & 84.868\% \\
    Precision & 92.308\% & 91.489\% \\
    $F_1$ Index & 0.932 & 0.881 \\
    $F_2$ Index & 0.938 & 0.861 \\
    $F_{0.5}$ Index & 0.927 & 0.901 \\
    LR+   & 1.020 & 0.990 \\
    Area under curve (AUC) & 0.963 & 0.931 \\
    \toprule
    The average number of B-lines &&\\
    \hspace{0.2cm} (Ground Truth) = 1.520 &&\\
    \hline
    Average Detected B-lines & 1.550 & 1.410 \\
    NMSE of number of & \multirow{2}[0]{*}{0.151} & \multirow{2}[0]{*}{0.243} \\
       \hspace{0.2cm} detected B-lines &       &  \\
       \toprule
    \end{tabularx}\vspace{-0.3cm}
  \label{tab:results}%
\end{table}%

Table \ref{tab:results} presents the B-line quantification results for the proposed method and PUI \cite{anantrasirichai2017line} in terms of the metrics given in Table \ref{tab:perfMetrics}, based on manual detections given by two clinical doctors experts in LUS. These quantitative results reveal that the proposed method improves B-line identification performance by 8\% when compared to PUI. Both methods had the same false detection performance whilst the missed detection performance of the proposed method is around 8\% less then PUI. The main important reason for this performance can be expressed via the proposed validation mechanism in the proposed methodology (Step 7), which is not performed in PUI. As illustrated in Fig. \ref{fig:step7}, our approach successfully discards non-B-line detections and reduces the percentage of missed detections by up to 5\%. Furthermore, $F_\beta$ metric clearly demonstrates that the proposed method is better for three different weighting performed for recall and precision metrics.

\begin{figure}[htbp]
\centering
\includegraphics[width=0.59\linewidth]{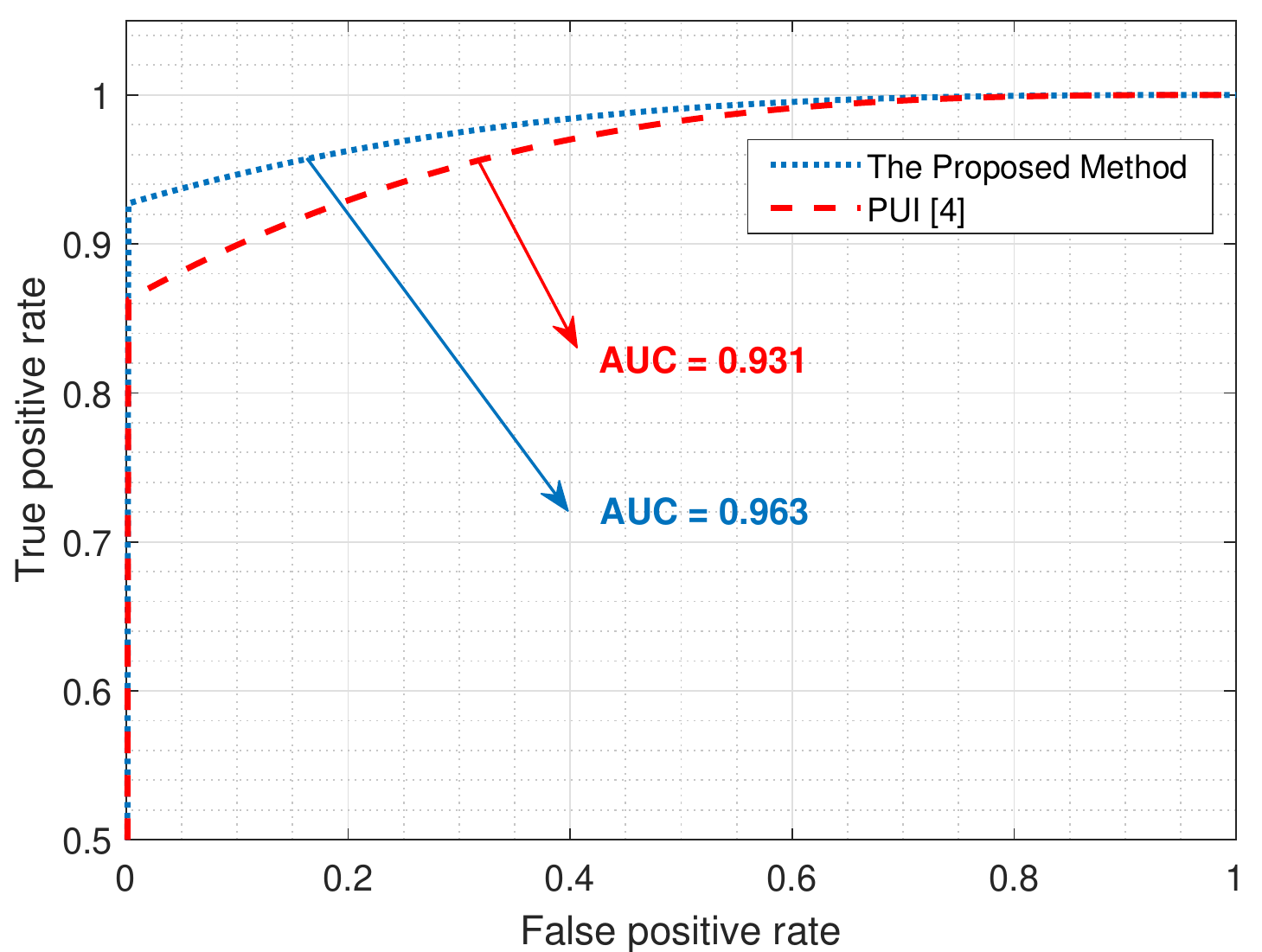}
\caption{Performance comparison of the B-line identification methods through a ROC curve.}
\label{fig:roc}\vspace{-0.3cm}
\end{figure}

Figure \ref{fig:roc} shows a receiver operating characteristic (ROC) curve, which illustrates the performances of the B-line identification methods via TPR and FPR. We considered the existence of B-lines and the non-existence of B-lines as positive and negative classes, respectively. The ROC curve shows that the proposed method significantly outperforms PUI. This also confirms the robust characteristics of the proposed approach since it achieves high recall (sensitivity) and area under curve (AUC) values.

\begin{figure*}[t]
\centering
\subfigure[]{\includegraphics[width=.19\linewidth]{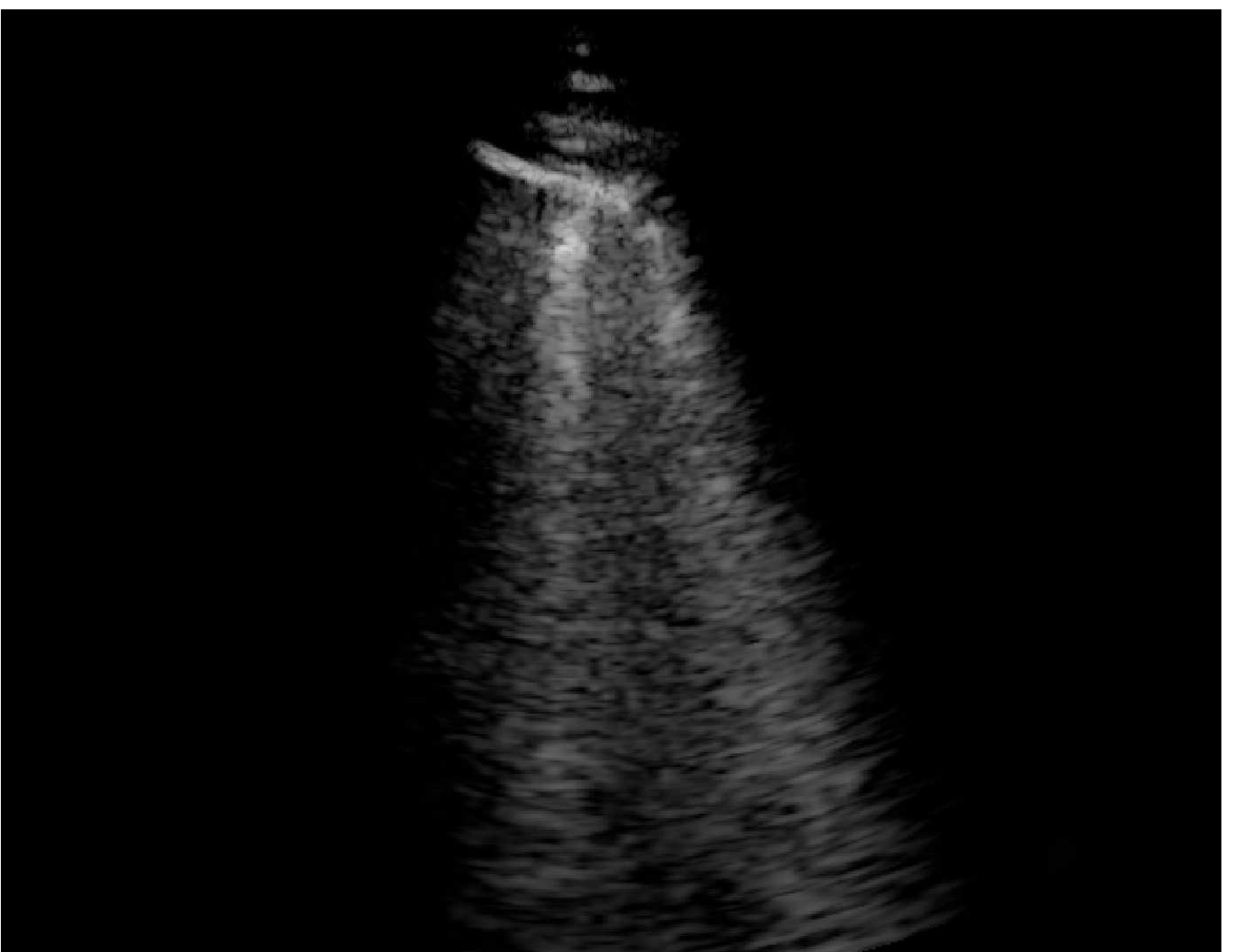}}
\subfigure[]{\includegraphics[width=.19\linewidth]{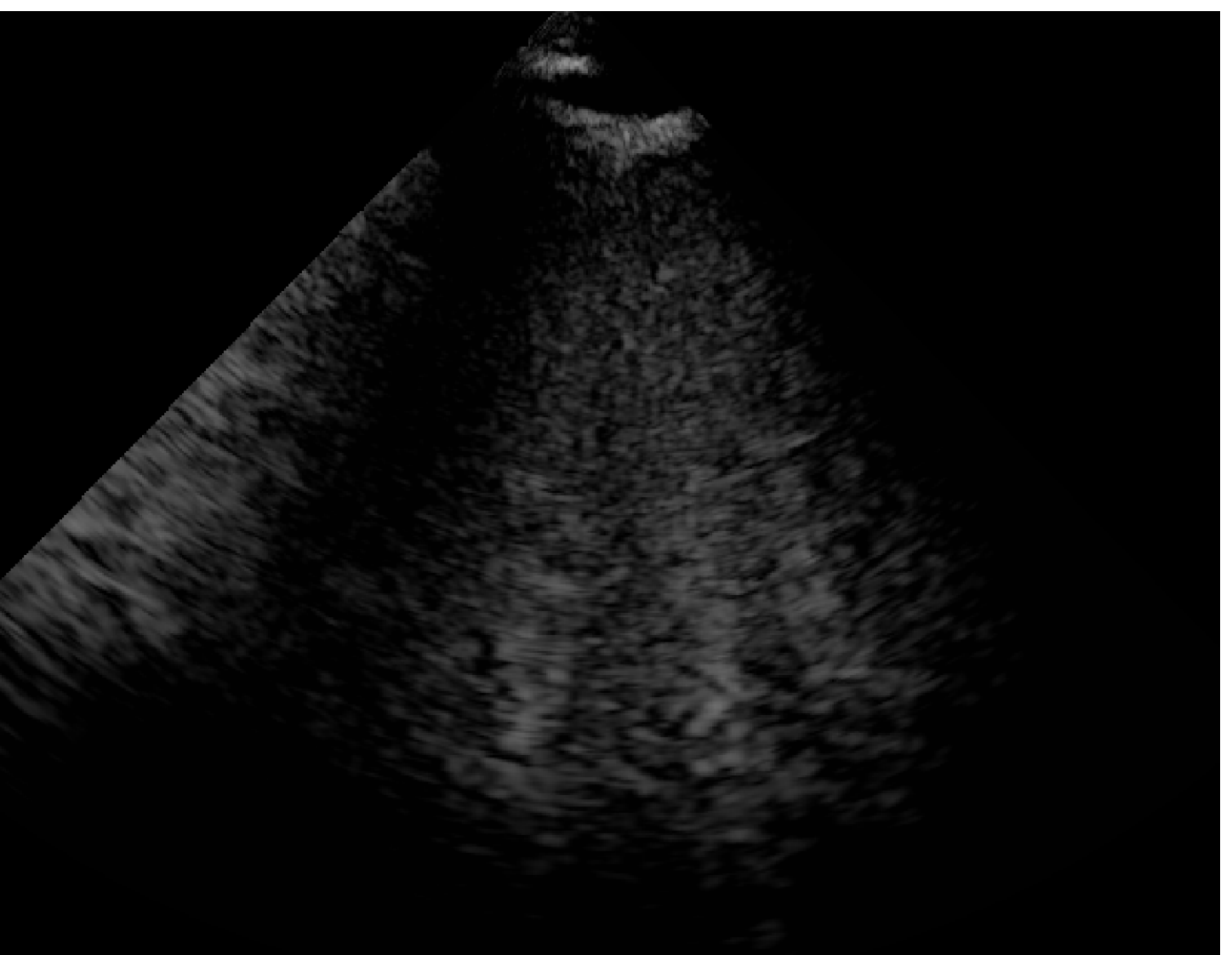}}
\subfigure[]{\includegraphics[width=.19\linewidth]{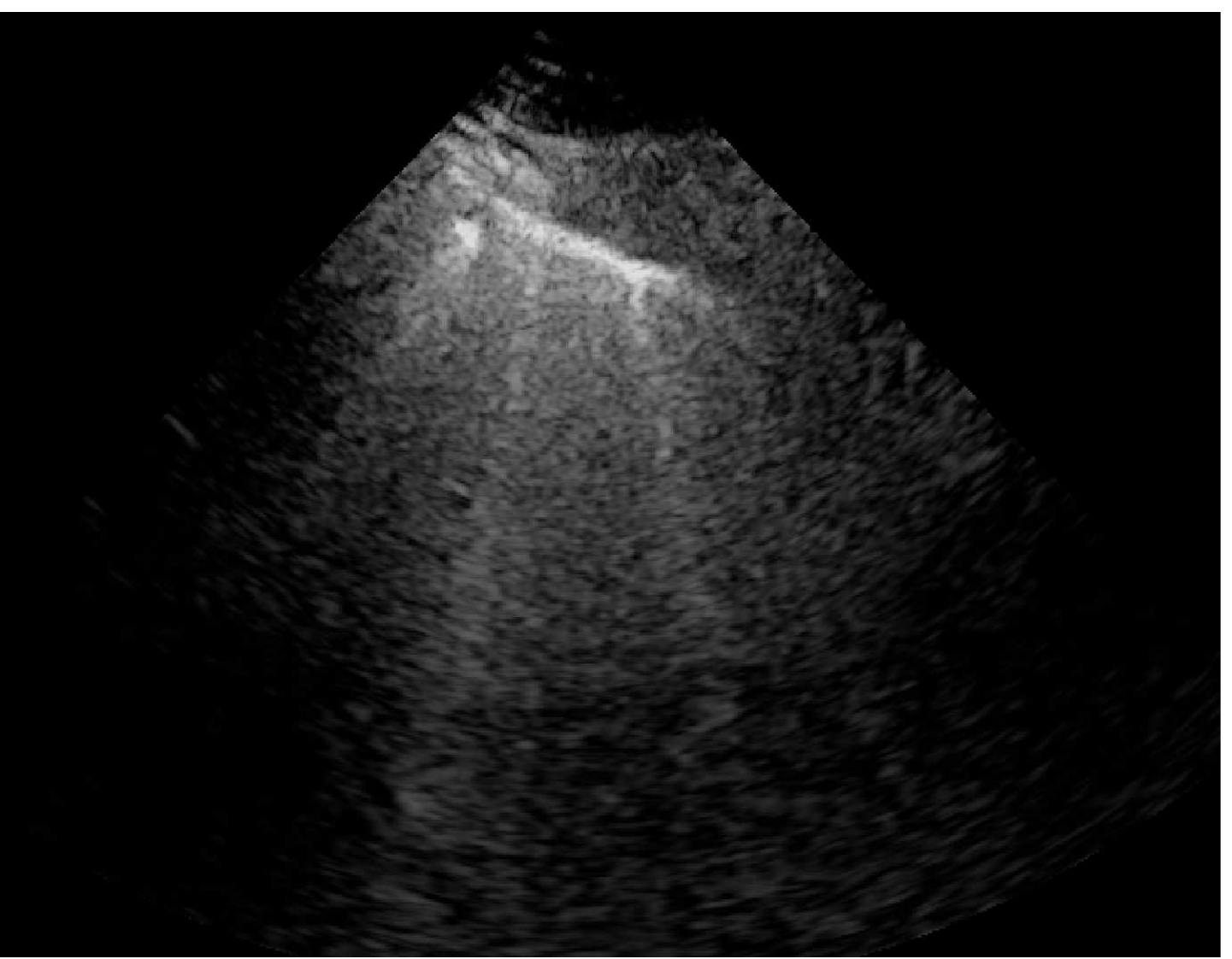}}
\subfigure[]{\includegraphics[width=.19\linewidth]{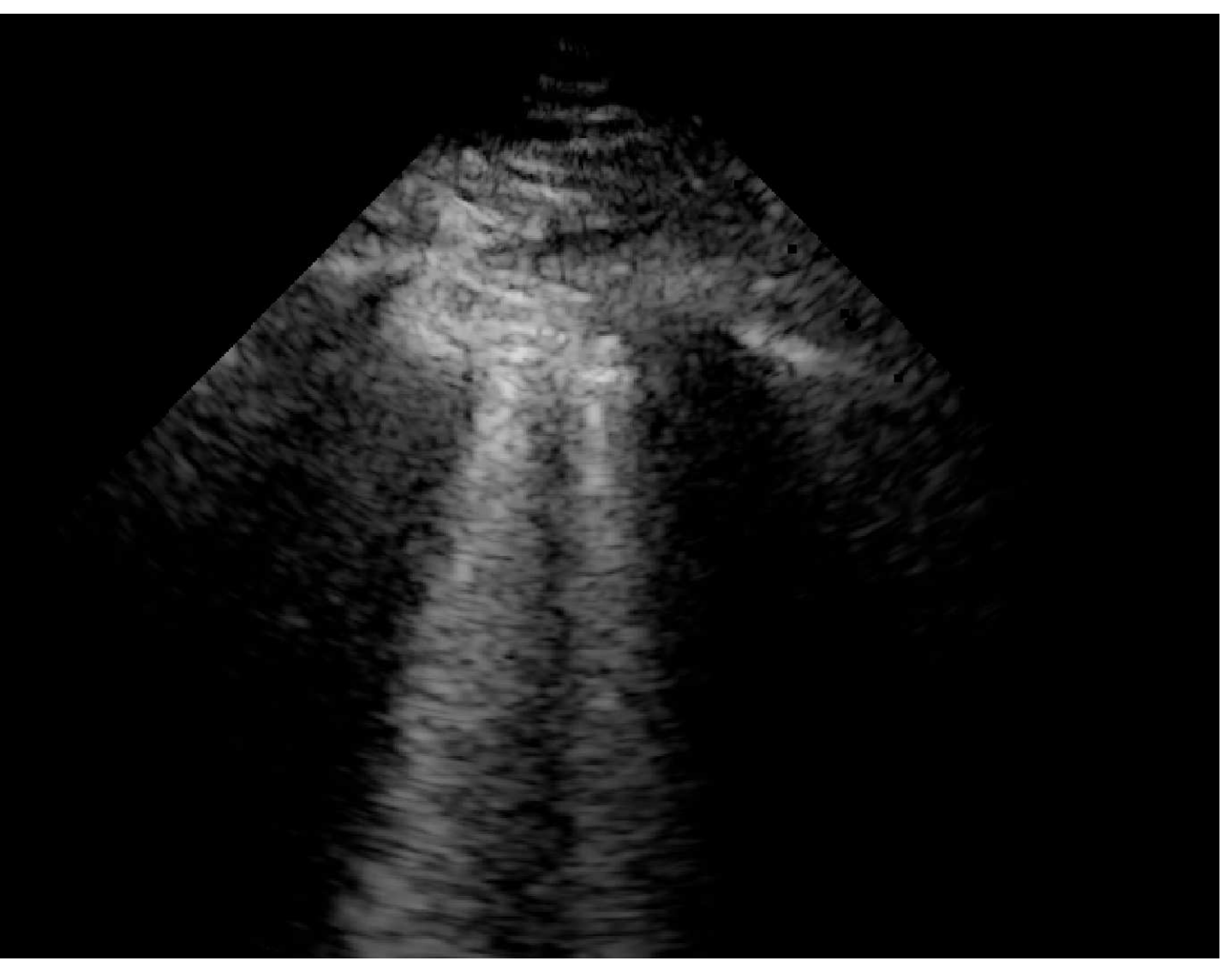}}
\subfigure[]{\includegraphics[width=.19\linewidth]{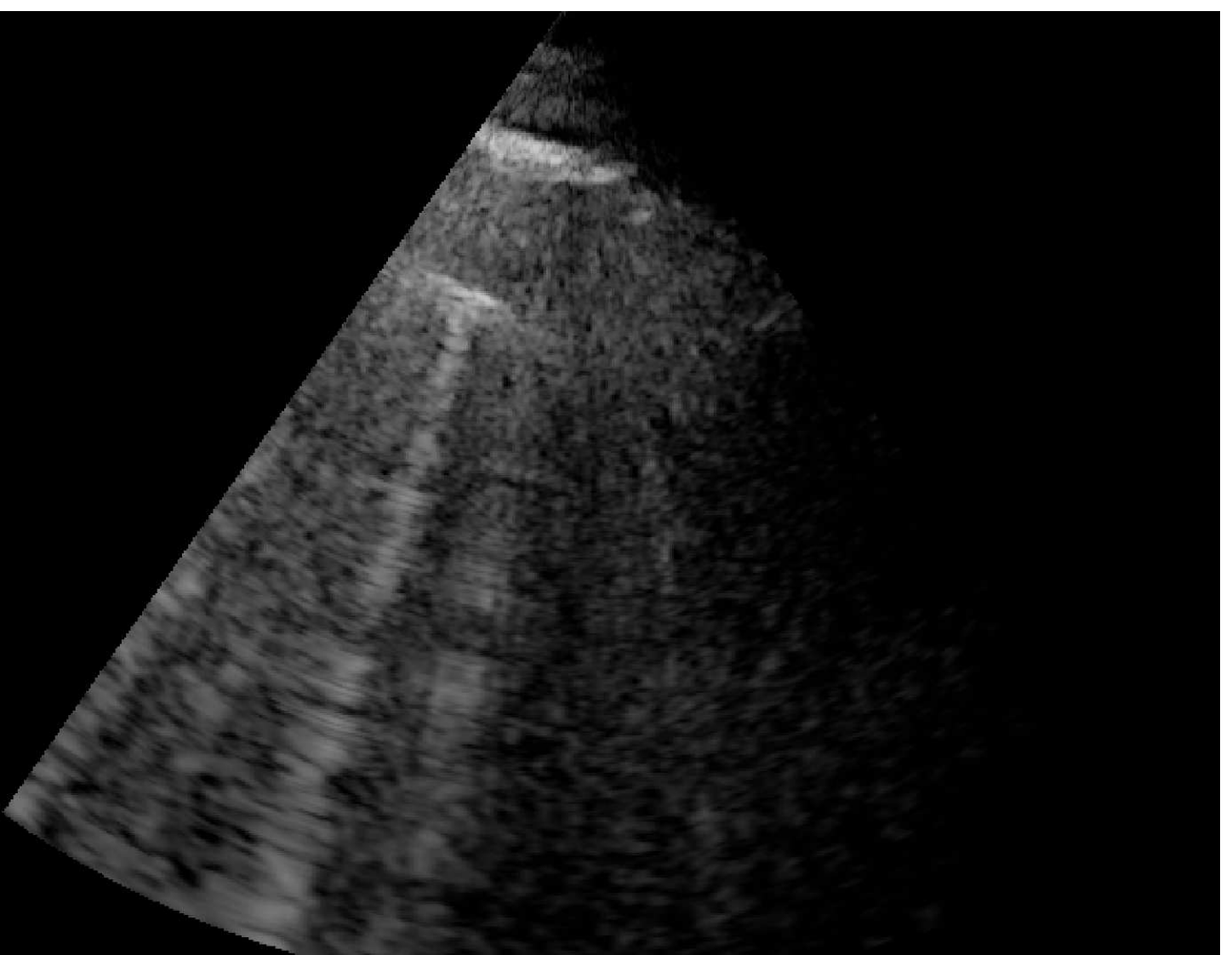}}
\subfigure[]{\includegraphics[width=.19\linewidth]{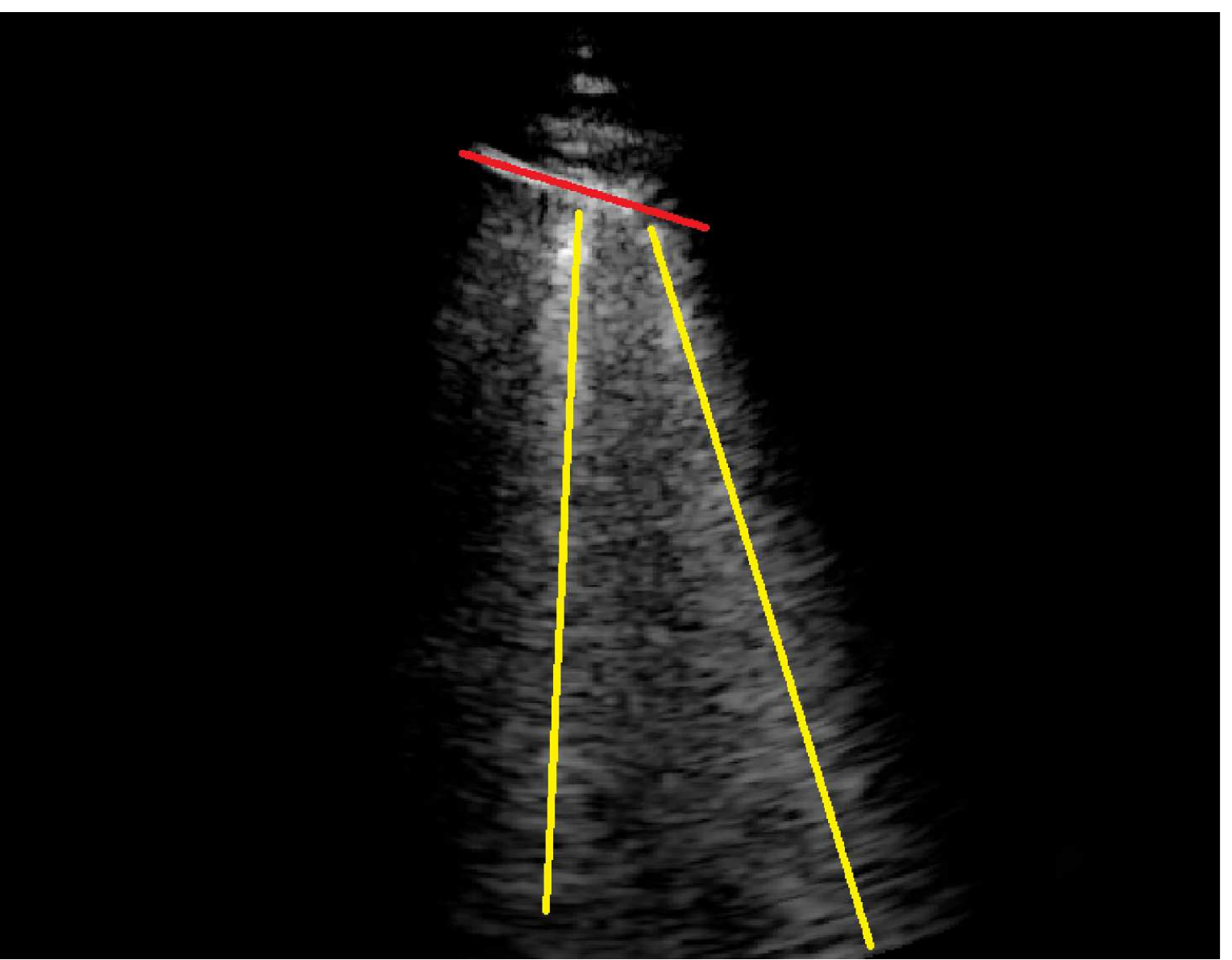}}
\subfigure[]{\includegraphics[width=.19\linewidth]{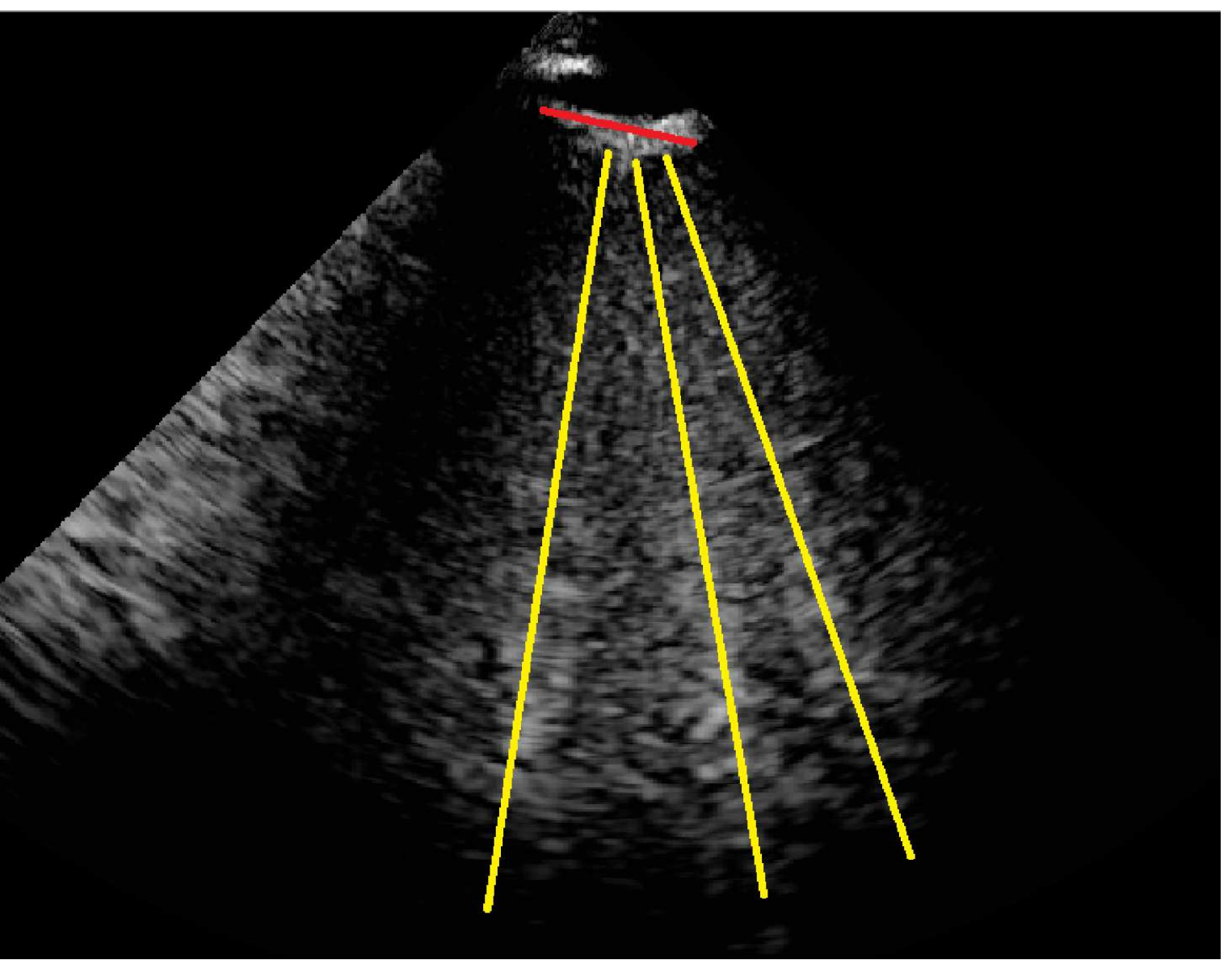}}
\subfigure[]{\includegraphics[width=.19\linewidth]{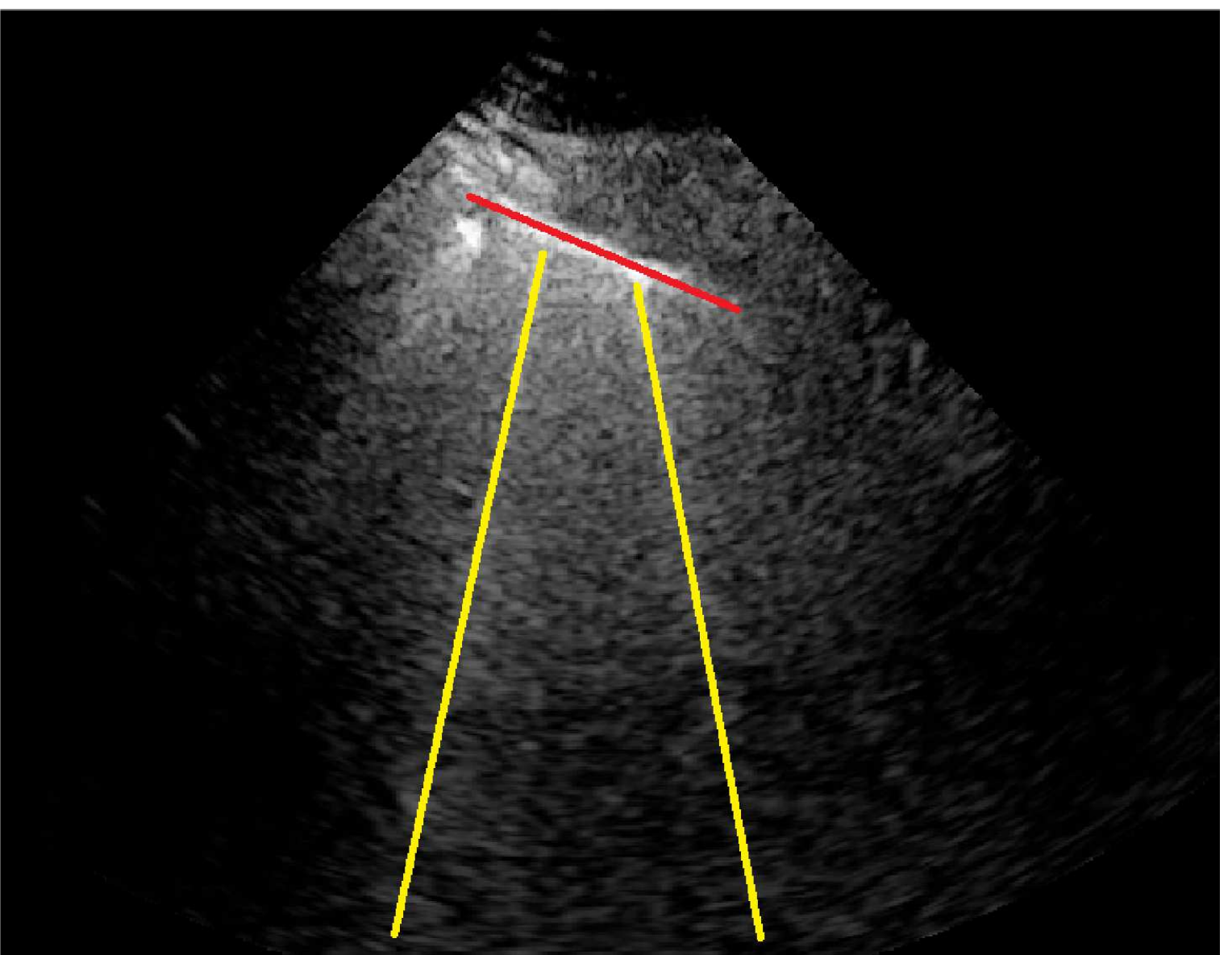}}
\subfigure[]{\includegraphics[width=.19\linewidth]{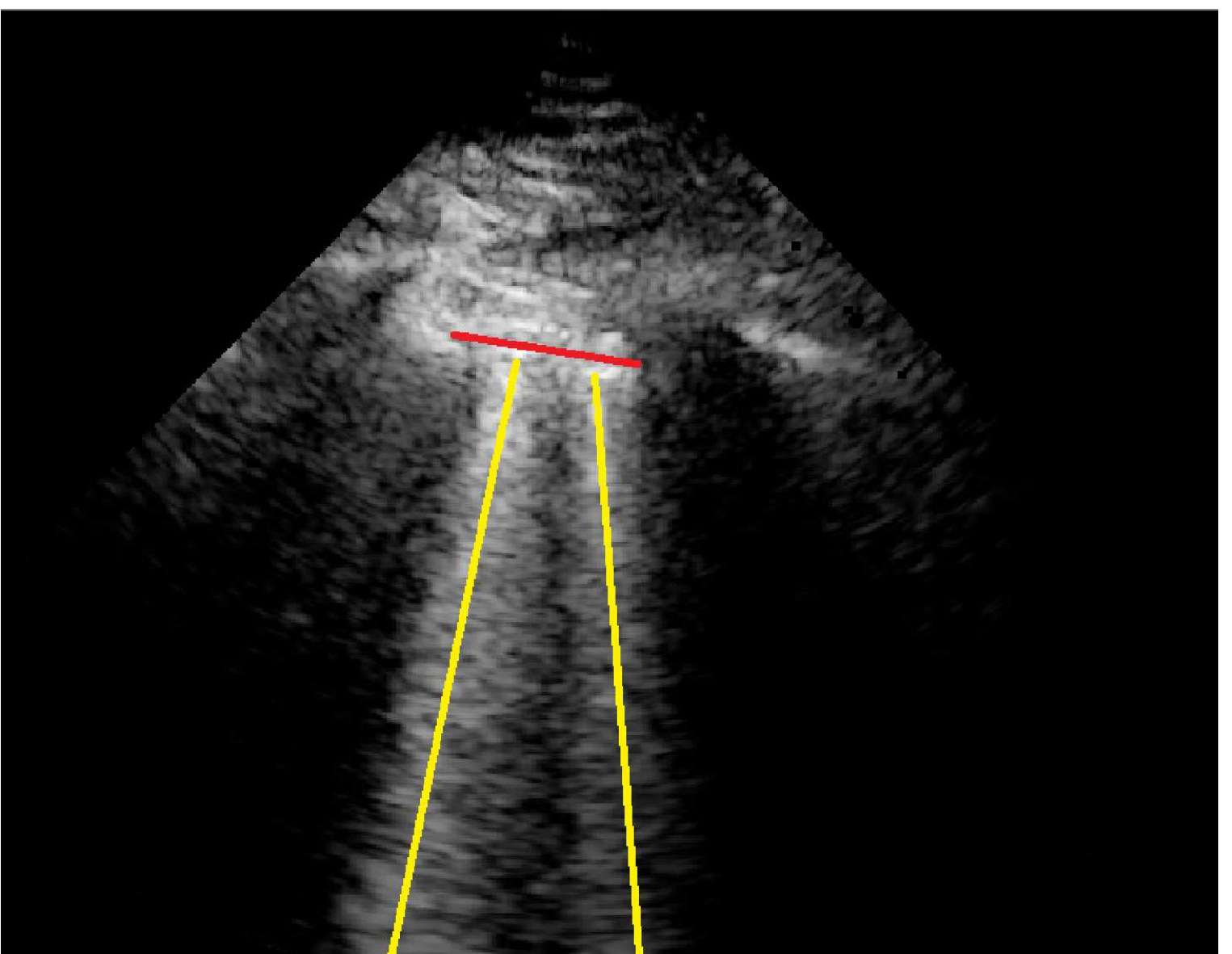}}
\subfigure[]{\includegraphics[width=.19\linewidth]{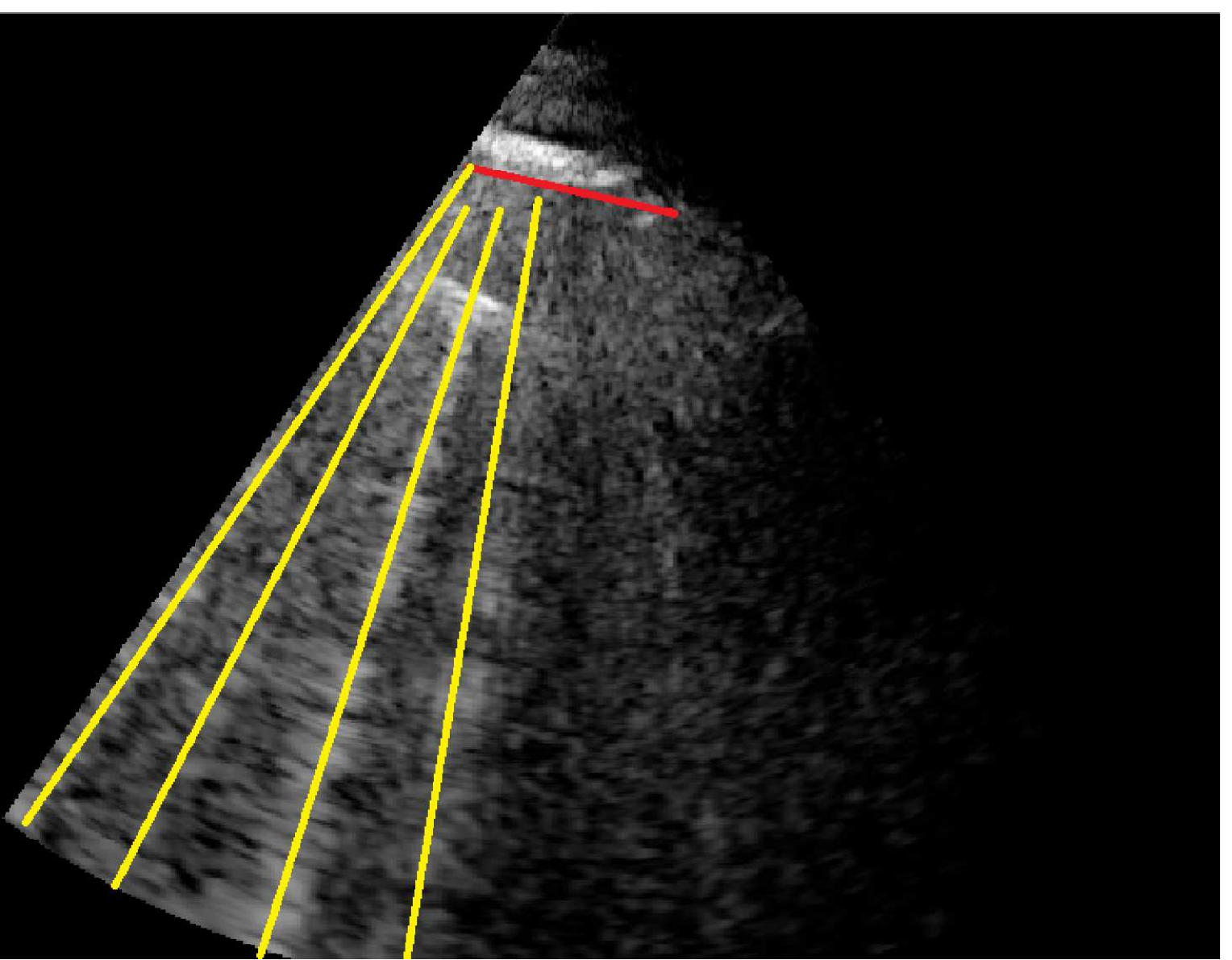}}
\subfigure[]{\includegraphics[width=.19\linewidth]{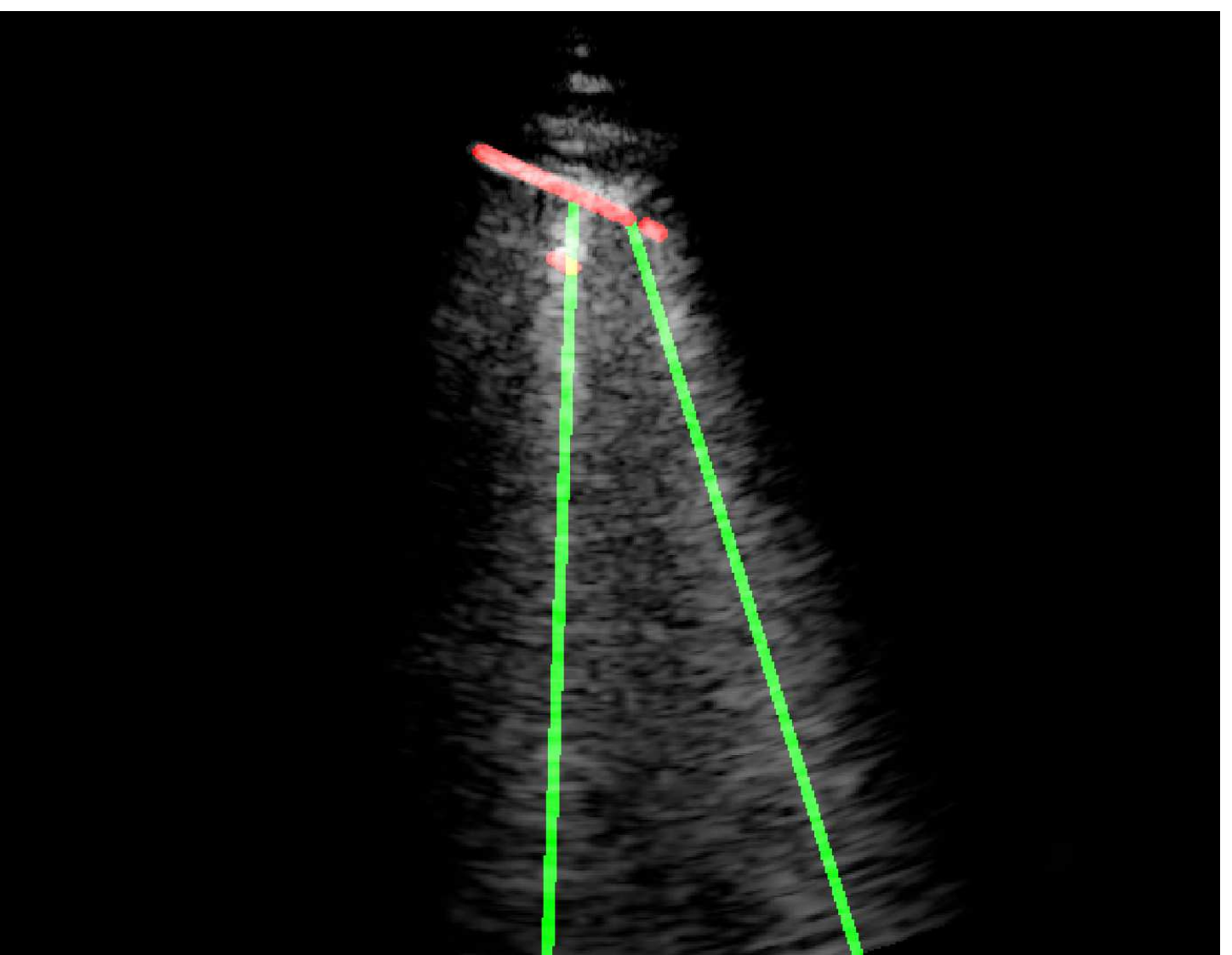}}
\subfigure[]{\includegraphics[width=.19\linewidth]{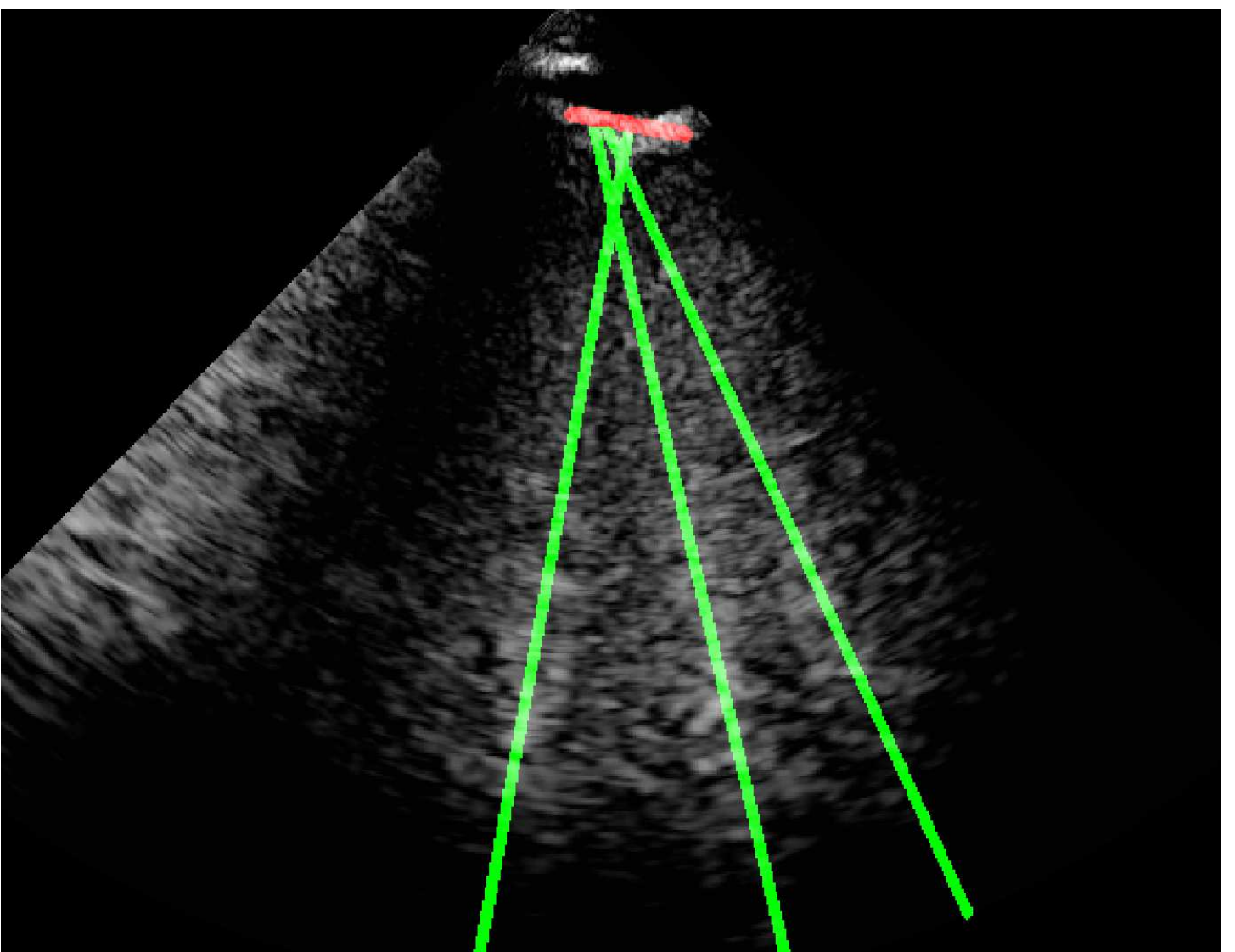}}
\subfigure[]{\includegraphics[width=.19\linewidth]{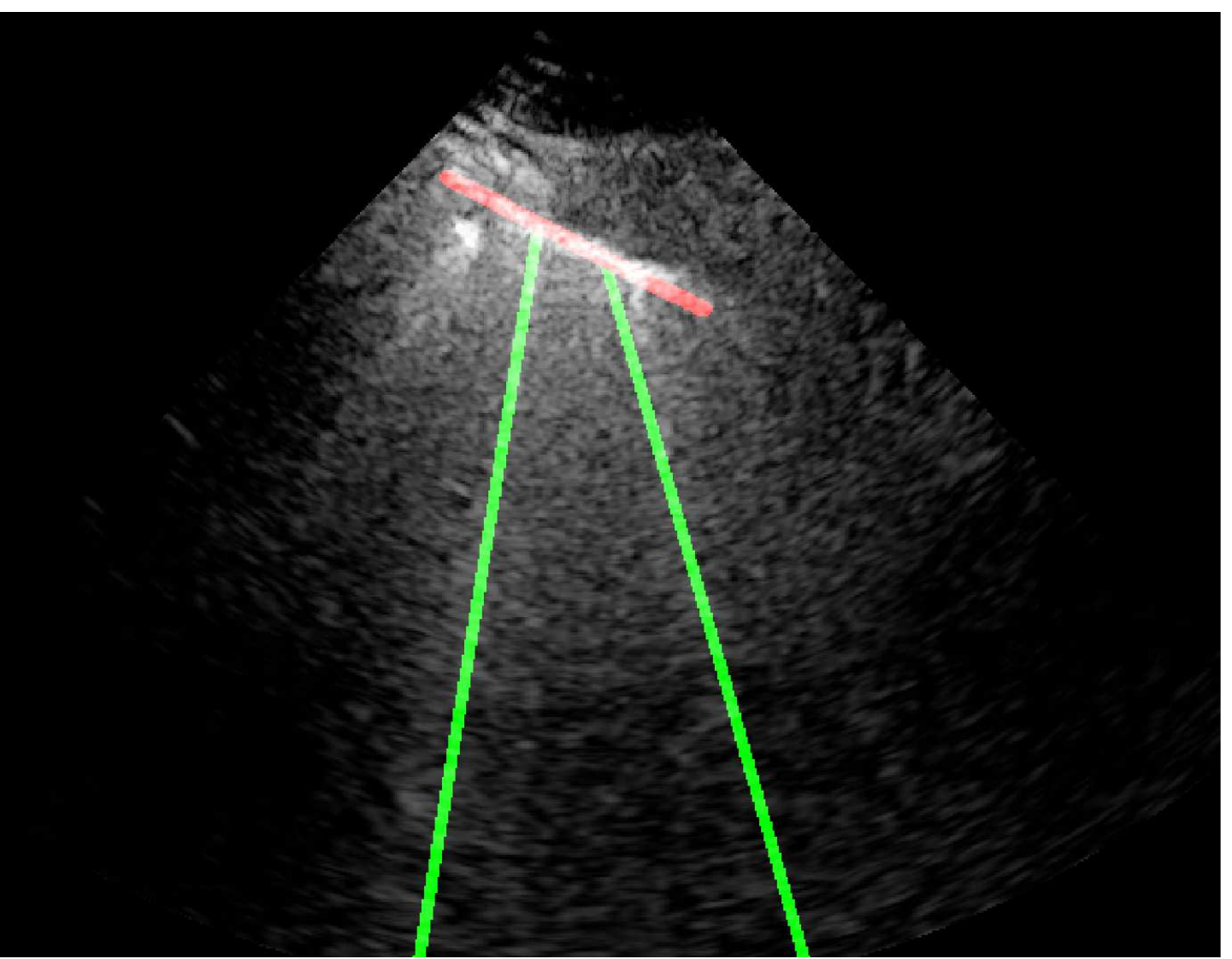}}
\subfigure[]{\includegraphics[width=.19\linewidth]{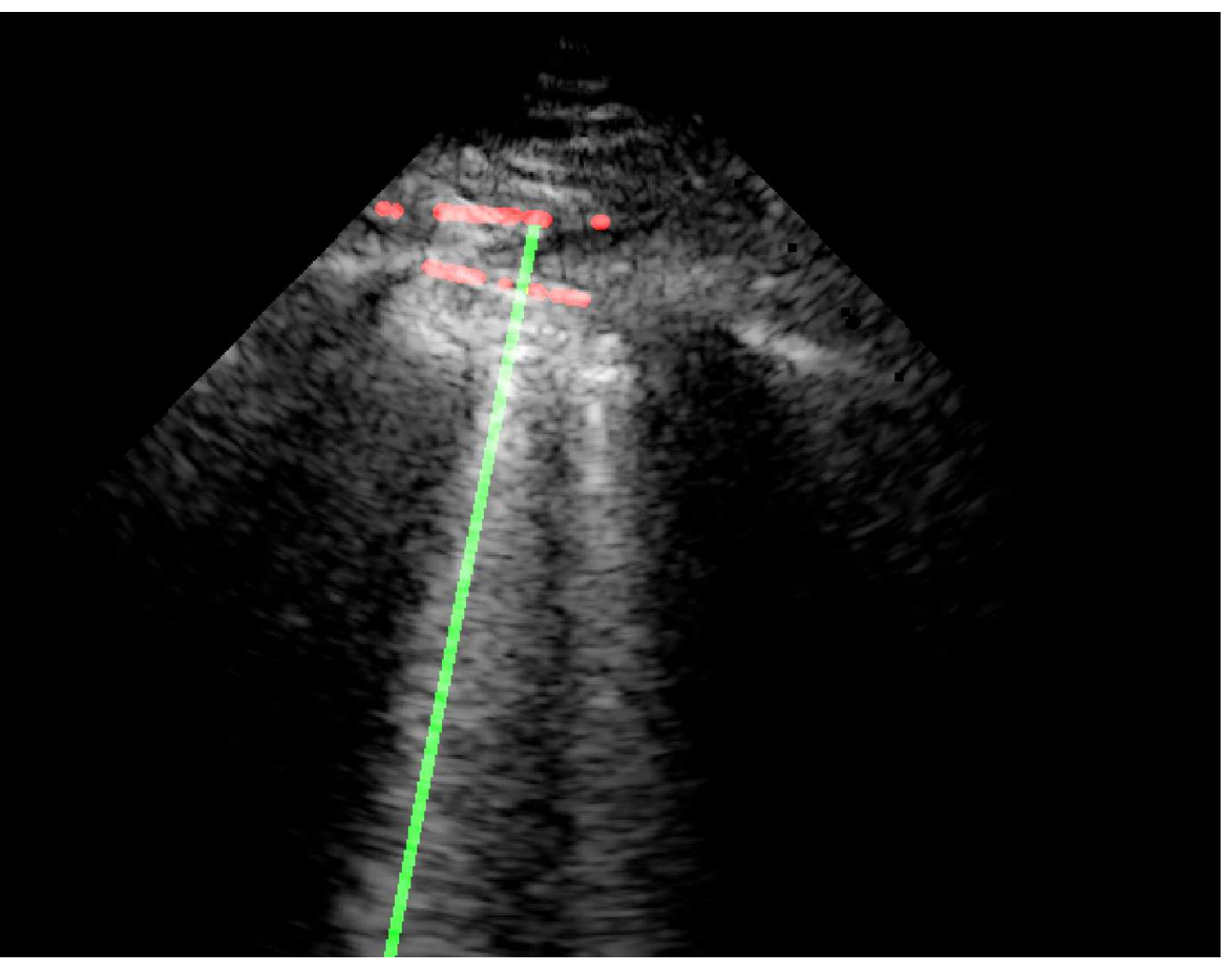}}
\subfigure[]{\includegraphics[width=.19\linewidth]{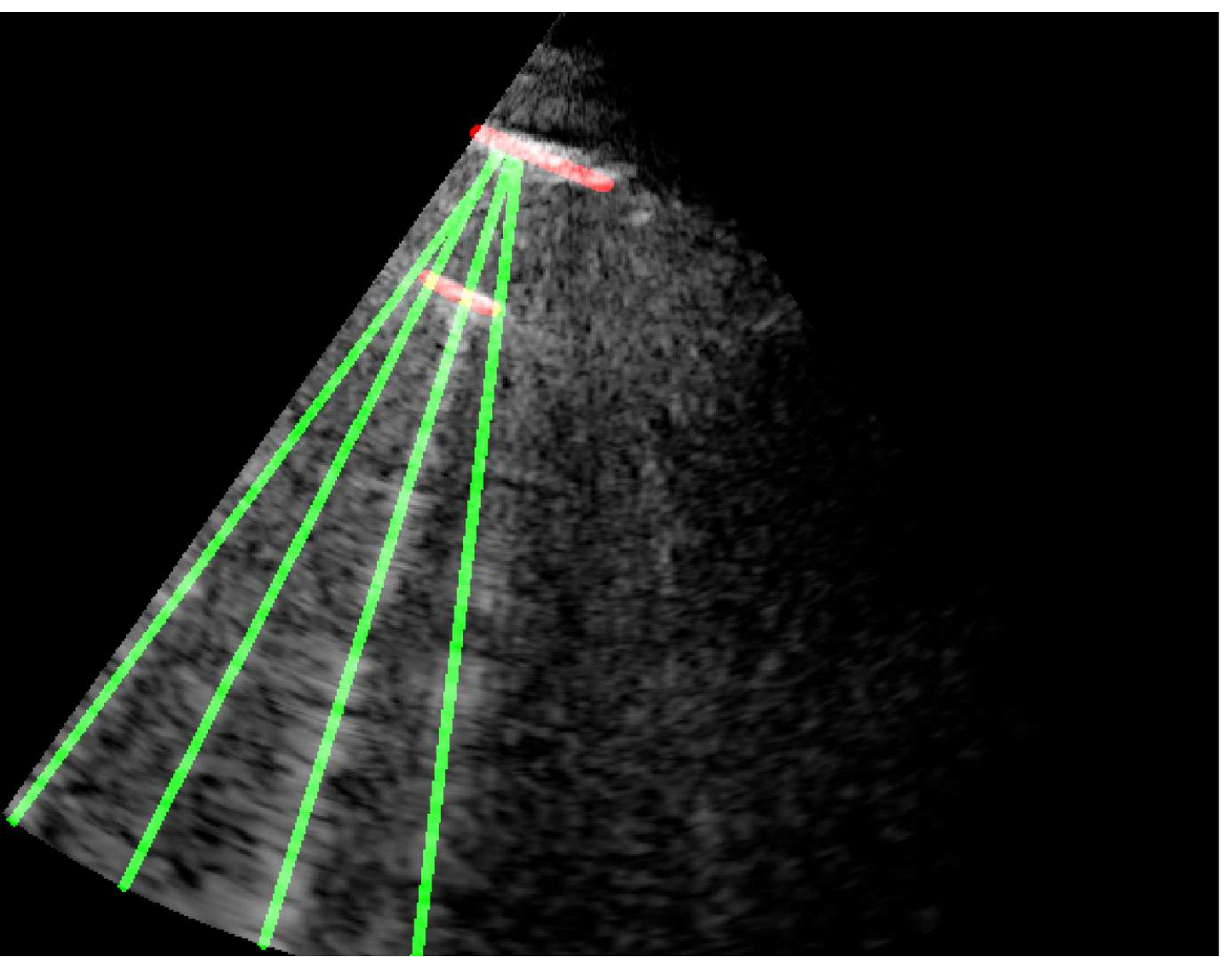}}
\subfigure[]{\includegraphics[width=.19\linewidth]{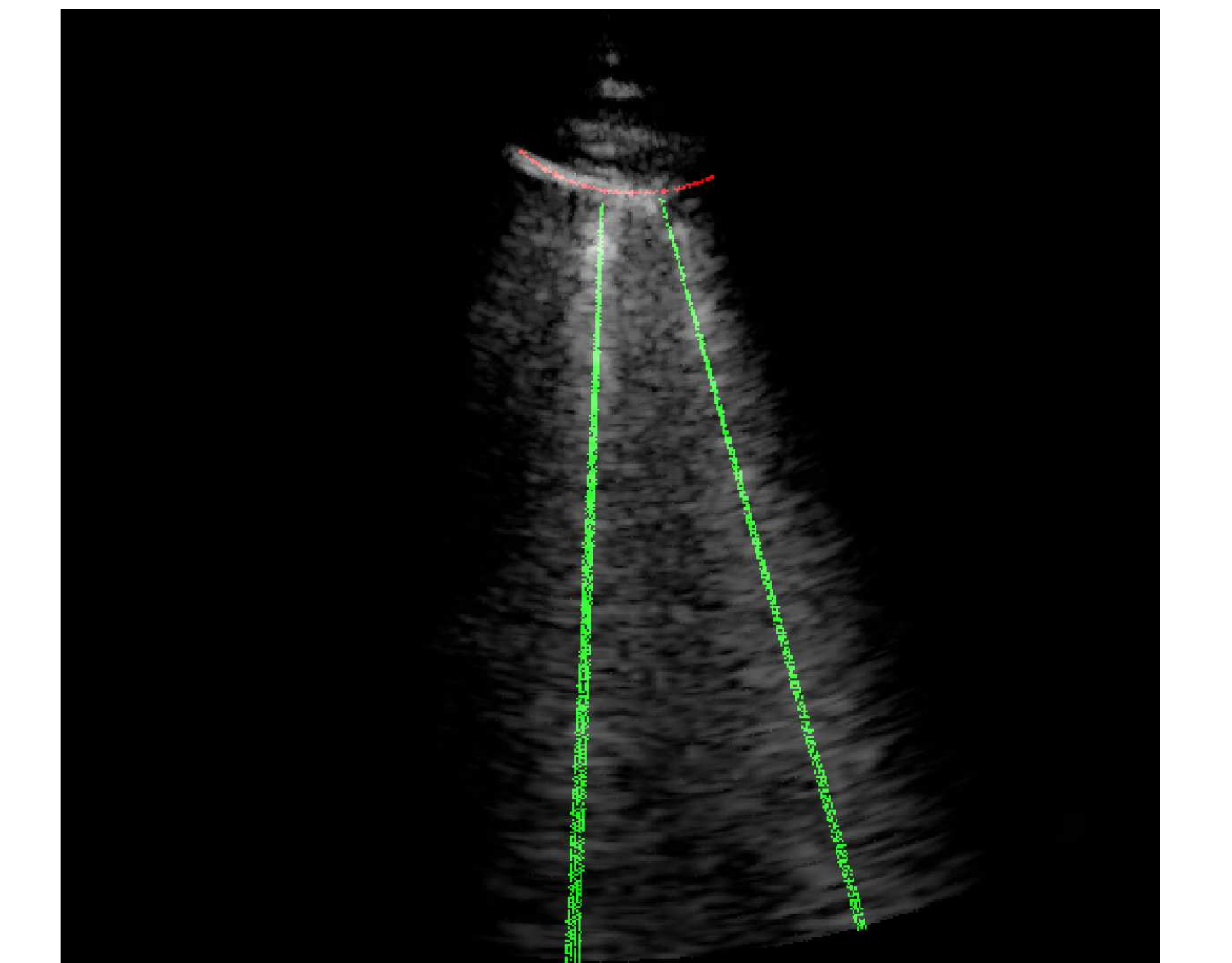}}
\subfigure[]{\includegraphics[width=.19\linewidth]{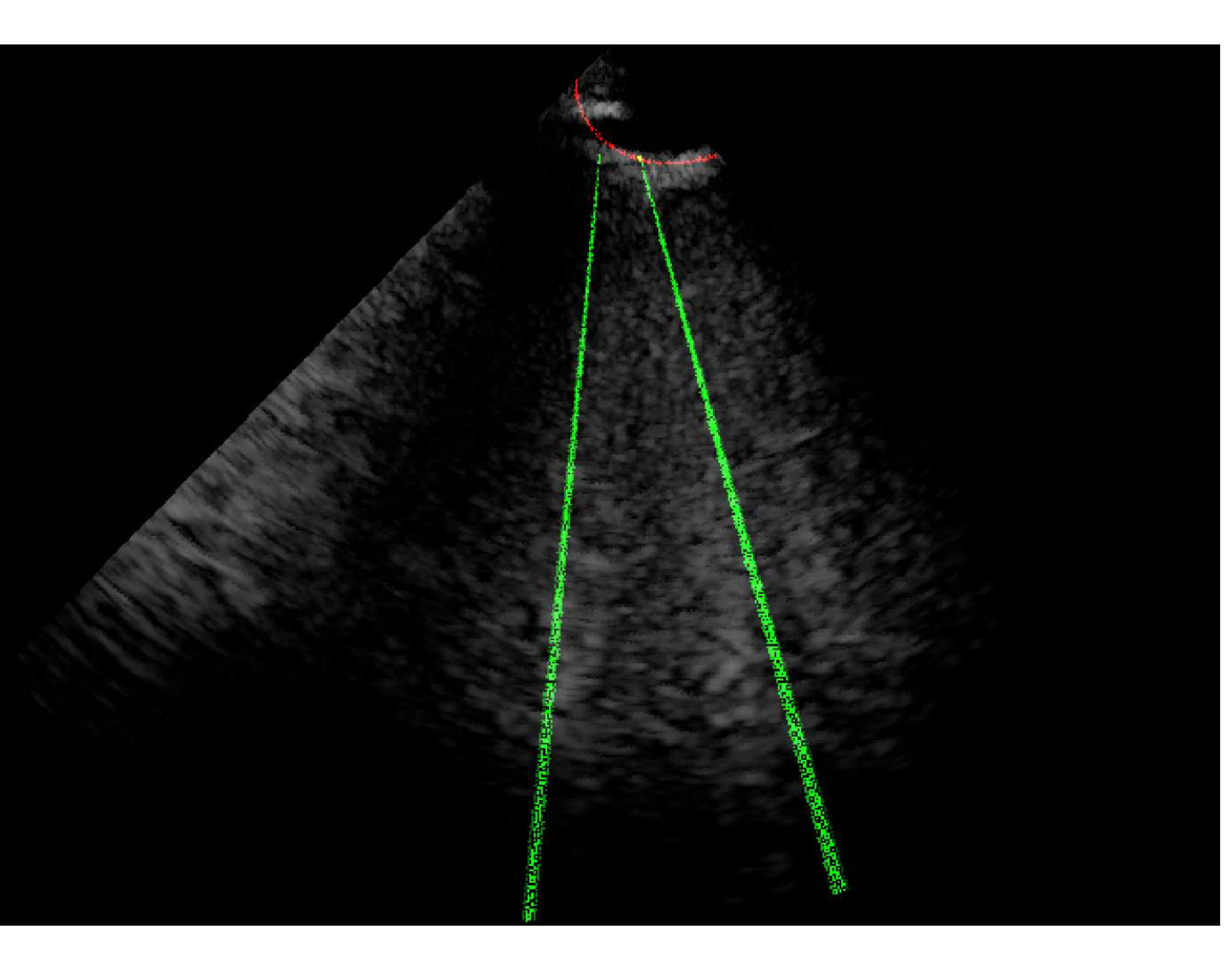}}
\subfigure[]{\includegraphics[width=.19\linewidth]{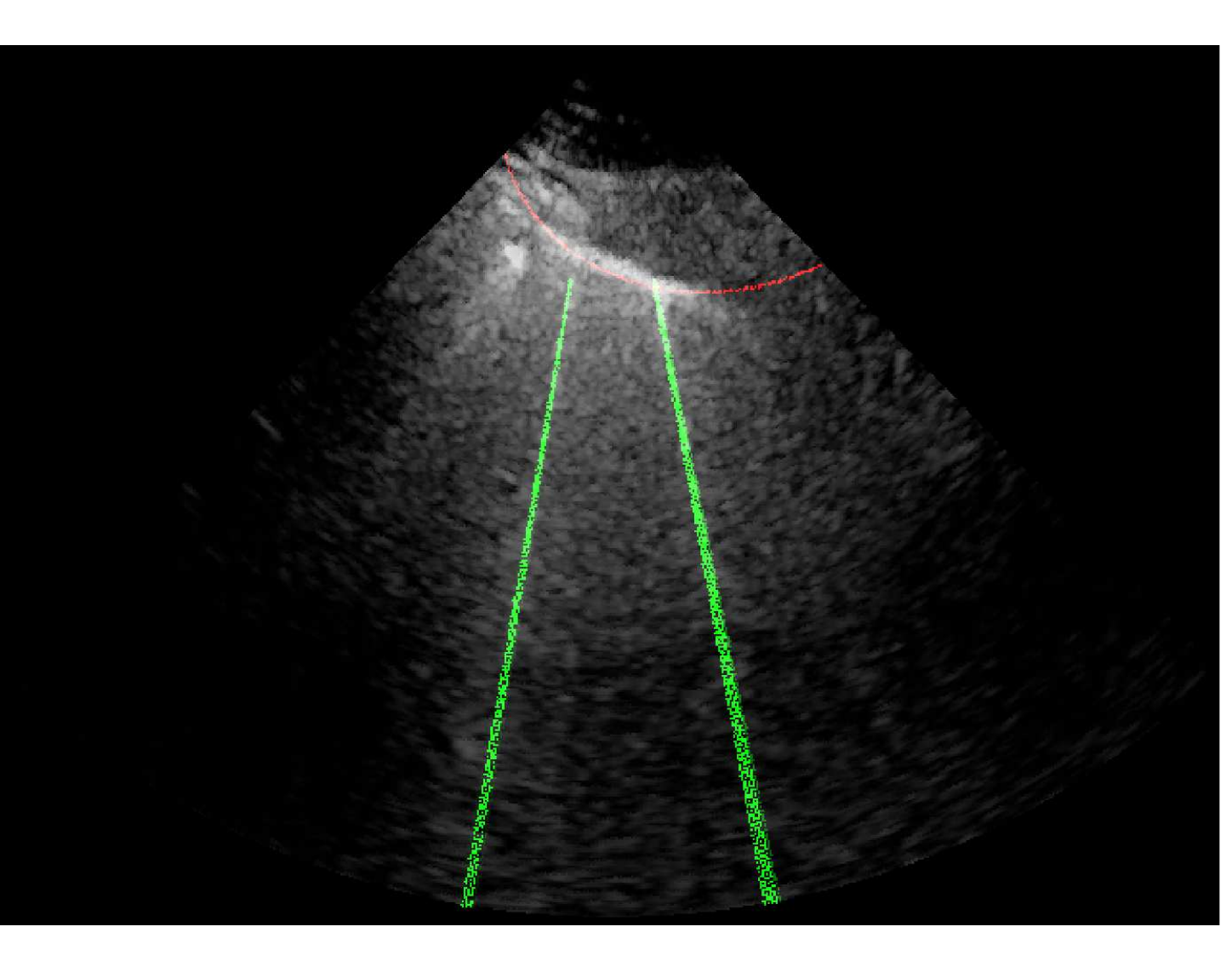}}
\subfigure[]{\includegraphics[width=.19\linewidth]{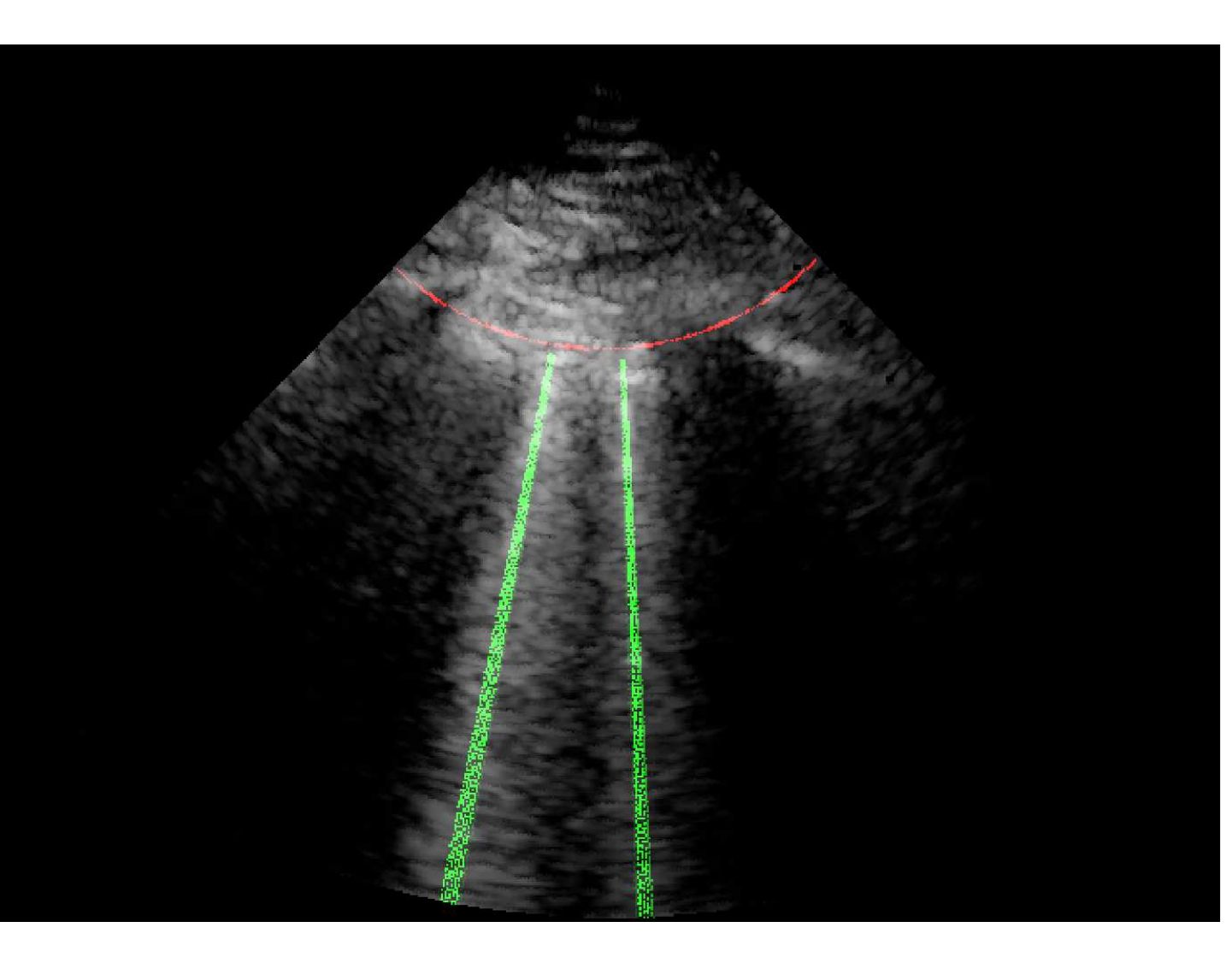}}
\subfigure[]{\includegraphics[width=.19\linewidth]{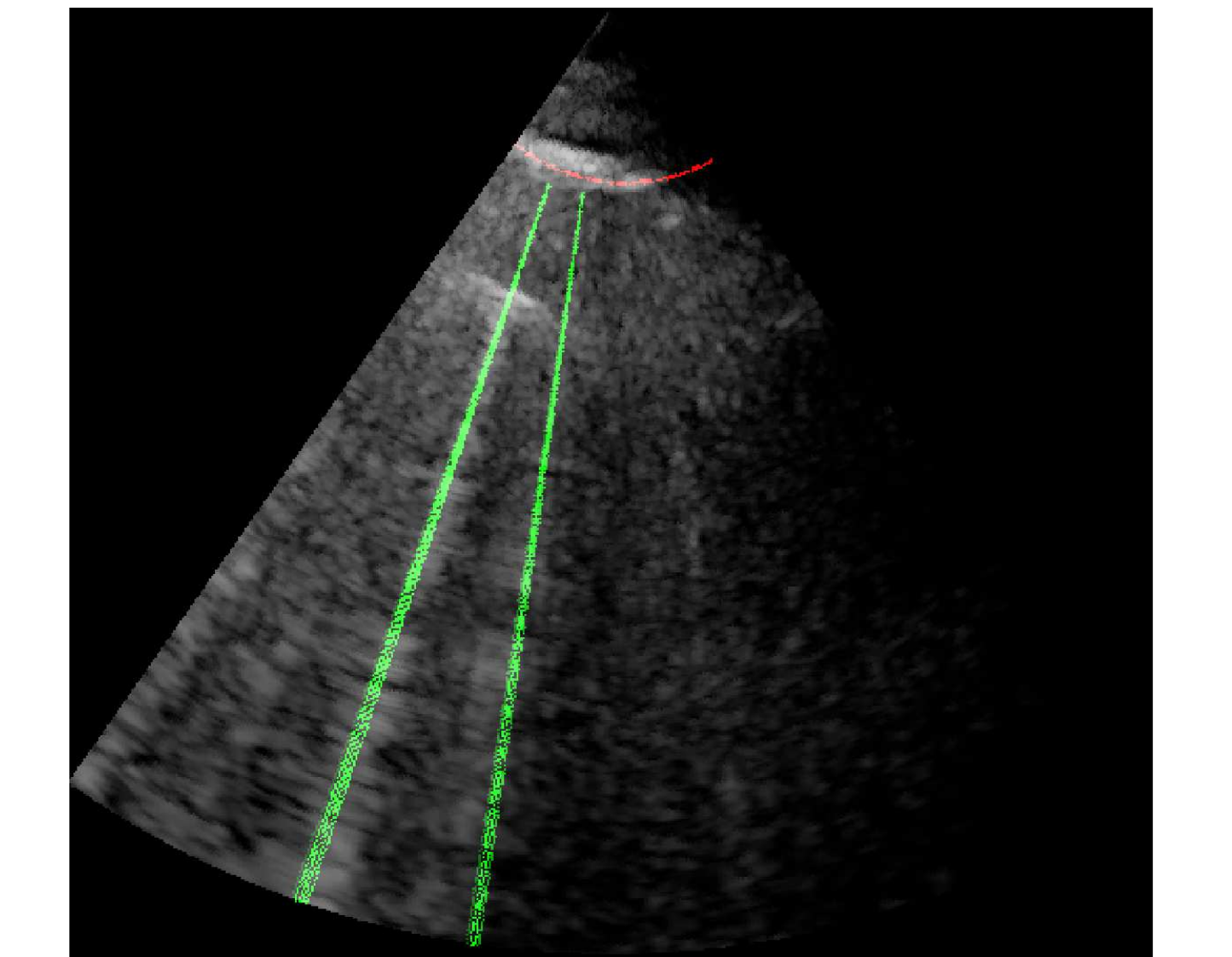}}
\caption{Detection results. (a)-(e) Original Images. (f)-(j) Ground truth. (k)-(o) The proposed method. (p)-(t) PUI  \cite{anantrasirichai2017line}. }\vspace{-0.3cm}
\label{fig:res1}
\end{figure*}

Figure \ref{fig:res1} shows the line artefact quantification results for various frames captured for different patients. The detected horizontal lines were drawn on the LUS images with red coloured lines, whilst vertical lines were drawn with green (yellow for ground truth annotations) coloured lines. Each row in Figure \ref{fig:res1} represents the original, the expert annotated detection results, and PUI results, respectively. When examining each example LUS image along with the detection results, we can state that for the LUS images in (a), (b), (c) and (e) the proposed method detects all correct B-lines and pleural lines correctly. The performance of PUI is better than the proposed method for LUS image in (d), whilst it includes few missed detections for LUS images in (b) and (e).

Another important metric shown in Table \ref{tab:results} is the average number of detected B-lines over all the 100 LUS images. One can observe that the proposed method's average B-line value is closer to the correct number of average B-lines with a normalised mean square error (NMSE) of 0.151, whilst PUI has an NMSE of 0.243. Among all 100 LUS images utilised in this study, the maximum number of B-lines in a single frame is 4 (Figure \ref{fig:res1}-(e)). The proposed method detects all of these correctly whereas PUI  achieves a maximum of 3 correctly detected B-lines in a single frame.

\begin{figure*}[ht]
\centering
\includegraphics[width=0.99\linewidth]{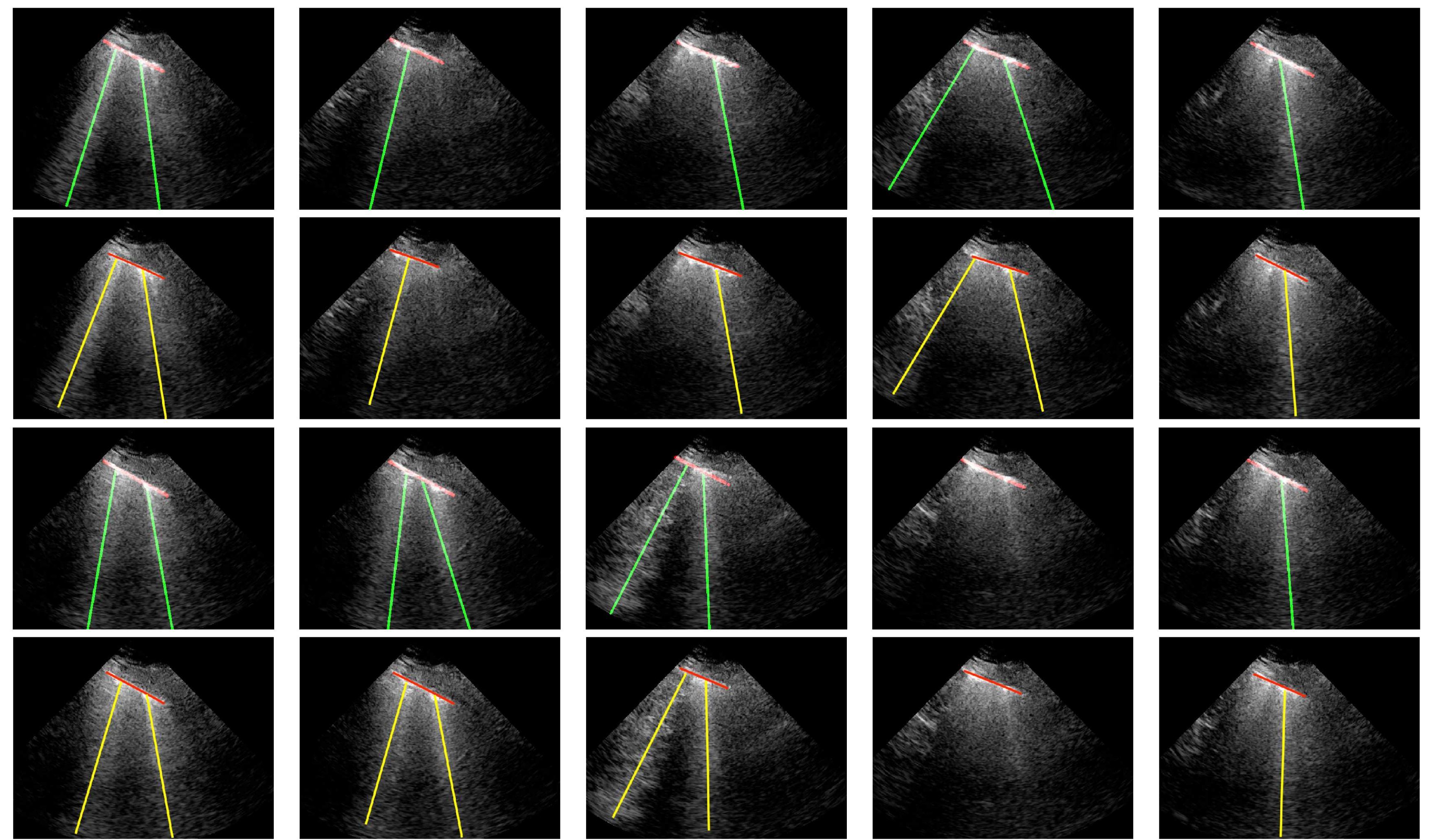}
\caption{The proposed image sequence processing results for a single patient, for selected 10 example frames. The 1$^{\text{st}}$ and 3$^{\text{rd}}$ rows refer to the proposed method results, whilst 2$^{\text{nd}}$ and 4$^{\text{th}}$ rows are ground truth annotated results by an expert. Specifically, from left to right at the first two rows, frames numbers are 35, 53, 82, 87, and 110 whilst the last two rows refer to frames 121, 160, 172, 186, and 201.}
\label{fig:video}
\end{figure*}

Finally, we tested the merits of the proposed method in processing LUS image sequences. Specifically, we fed the algorithm with a sequence of frames (201 frames for each sequence), the line detection method was run on a frame by frame basis, and an image sequence was generated including the detected line artefacts. For this test, we used 5 out of 9 patients. Specifically, for each patient, we used a single lung region measurement. An example detection result for ten randomly selected LUS frames is given in Fig. \ref{fig:video} along with the expert annotated ground truth results. When examining the results, we can clearly see that for all ten randomly selected frames, the proposed method detects B-lines and pleural lines with 100\% accuracy.

\begin{figure*}[htbp]
\centering
\subfigure[]{\includegraphics[width=0.48\linewidth]{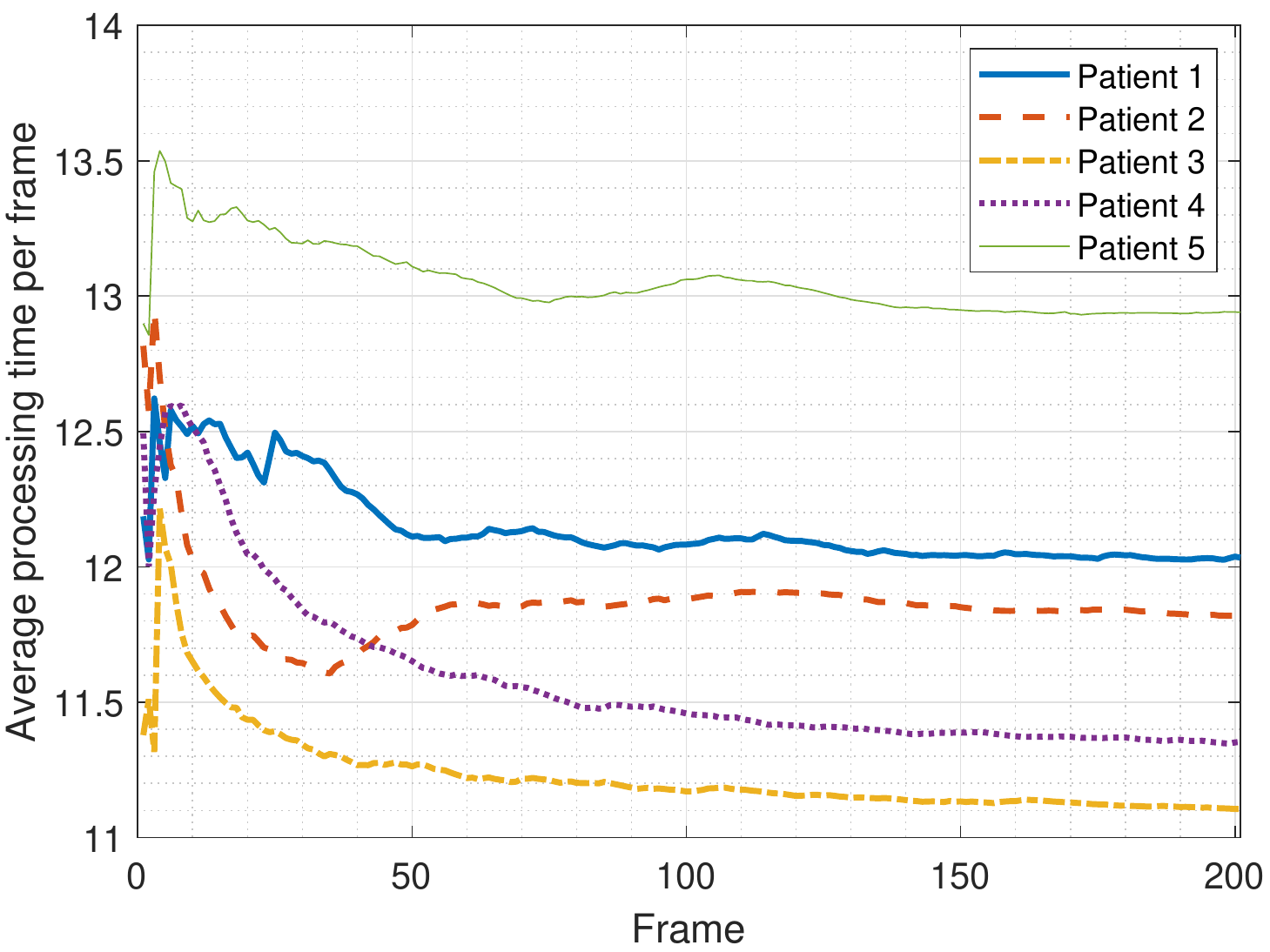}}
\subfigure[]{\includegraphics[width=0.48\linewidth]{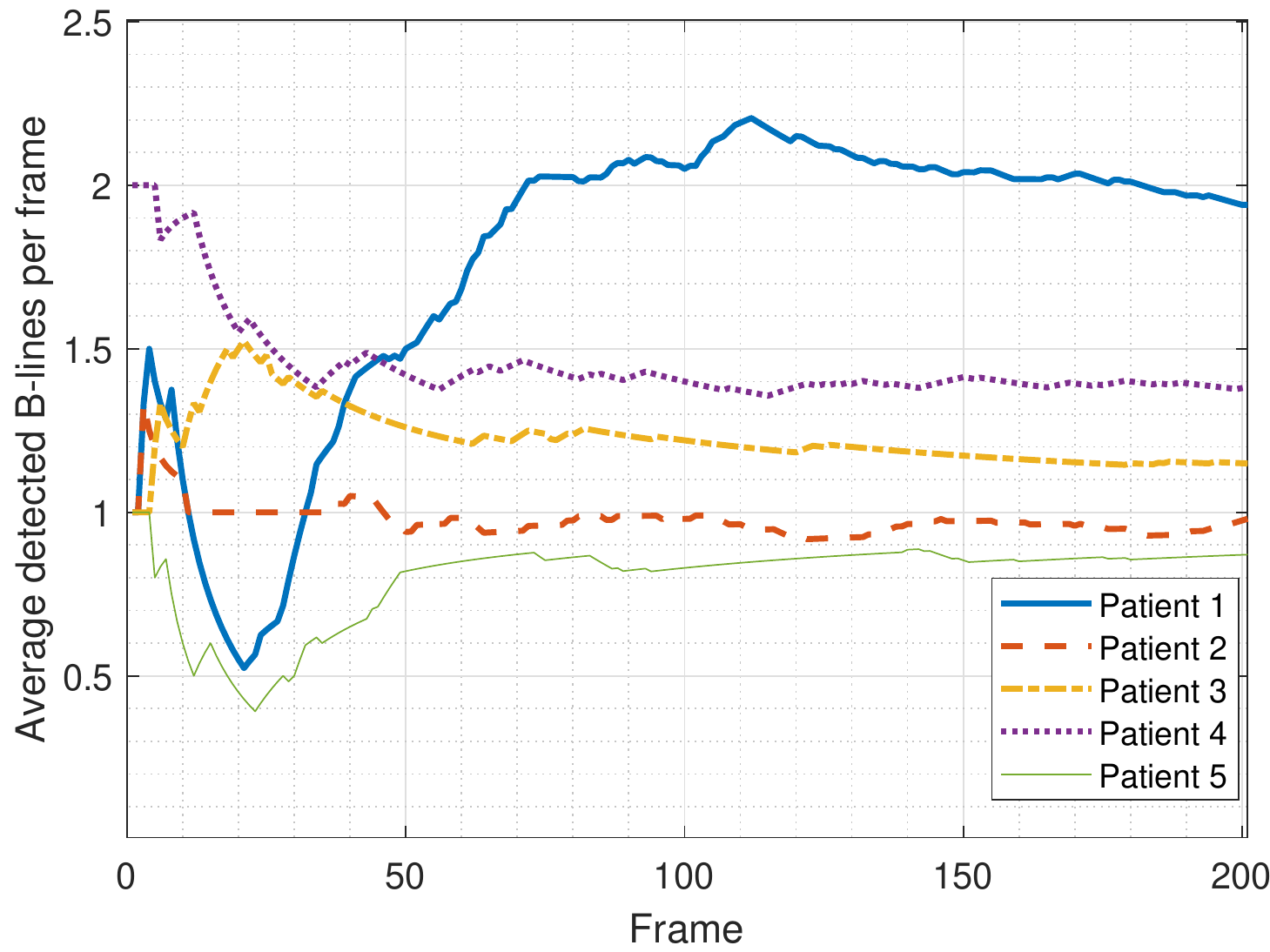}}
\caption{Image sequence processing results. Running mean plots for (a) processing time per frame, and (b) detected number of B-lines.}
\label{fig:video2}
\end{figure*}

Despite being based on the optimisation of a non-convex cost function, the advantage of the proposed methodology in processing image sequences is its guaranteed convergence, which is achieved with a reasonably low computational cost. In Fig. \ref{fig:video2}-(a), we show the mean processing time per frame for all five image sequences we utilised in this study. On examining Fig. \ref{fig:video2}-(a), one can clearly see that the processing time for a single frame is around 11-13 seconds. The method is implemented in Matlab R2019a on a laptop computer (Intel i7-2.7 GHz processor with 16 GB RAM). Note that the relatively long computational time could be hugely reduced following code optimization and eventual parallelization.

Figure \ref{fig:video2}-(b) shows the mean of the detected B-lines per frame for all five patients. From this figure, one can see that the average number of B-lines for different COVID-19 patients is generally around 1-1.5, which is correlated with the average number of correct B-lines (1.520) over all 100 LUS images employed in the previous experiment.

\section{Conclusion}\label{sec:conc}
Along with other medical imaging modalities, including CT and X-ray, lung ultrasound imaging has played an ever increasing role in confirming positive COVID-19 patients. Indeed, due to its applicability at the bedside and real time capability in assessing lungs status, LUS has quickly become a modality of choice during the SARS-CoV-2 outbreak. Line artefacts in LUS provide vital information on the stage and progression of COVID-19. Hence, automating the detection of B-lines will widen the applicability of LUS whilst reducing the need for expert interpretation and will benefit doctors, nurses, patients and their families alike.

In this paper, we proposed a novel non-convex regularisation based line artefacts quantification method with applications in LUS evaluation of COVID-19 patients. The proposed method poses line detection as an inverse problem and exploits Radon space information for promoting linear structures. We utilised a non-convex Cauchy-based penalty function, which guarantees convergence to global minima when using the Cauchy proximal splitting algorithm.

Experimental results demonstrate an excellent line detection performance for 100 LUS images showing several B-line structures, observed in nine COVID-19 patients. The performance evaluation of the proposed algorithm was conducted in comparison to the state-of-the-art PUI method in \cite{anantrasirichai2017line}. Objective results demonstrate that the proposed method outperforms PUI by around 8\% in terms of B-line detection accuracy. It is also important to note that the proposed method is completely unsupervised. On the assumption that annotated data is not available freely, an unsupervised method such as the proposed one will always have advantages over (weakly, semi- or fully) supervised methods.

Despite our current un-optimised Matlab implementation, the proven convergence properties of the proposed algorithm enable the computation of a single LUS frame in around 12 seconds. This determines the processing of LUS image sequences in a reasonable amount of time with a very good performance. Further optimisation of our algorithms for achieving real-time performance and their implementation on mobile platforms, and further evaluations in other acquisition settings are part of our current research endeavours.

\section*{Acknowledgment}
The authors would like to thank Dr Hélène Vinour, Dr Sihem Bouharaoua and Dr Béatrice Riu for help with data collection. They would also like to thank Prof. Olivier Lairez, from Toulouse University Hospital, who facilitated the initiation of this collaborative effort.

\bibliographystyle{IEEEtran}
\bibliography{COVID19_Blines}










\end{document}